\documentclass[apj,revtex4]{emulateapj}

\usepackage{amsmath}
\usepackage{lscape}
\usepackage{color}
\usepackage{hyperref}

\newcommand{\Msolar}{M$_{\odot}\,$}
\newcommand{\kms}{km s$^{-1}$}

\begin{document}

\title{On the Origin of Sub-subgiant Stars. I. Demographics}
\shorttitle{Demographics of Sub-subgiant Stars}

\author{Aaron M.\ Geller$^{1,2,\dagger,}$\footnote{NSF Astronomy and Astrophysics Postdoctoral Fellow}, 
Emily M.\ Leiner$^3$,
Andrea Bellini$^4$,
Robert Gleisinger$^{5,6}$,
Daryl Haggard$^5$,
Sebastian Kamann$^7$,
Nathan W.\ C.\ Leigh$^8$,
Robert D.\ Mathieu$^3$,
Alison Sills$^9$,
Laura L.\ Watkins$^4$,
David Zurek$^{8,10}$}

\affil{$^1$Center for Interdisciplinary Exploration and Research in Astrophysics (CIERA) and Department of Physics \& Astronomy, Northwestern University, 2145 Sheridan Rd., Evanston, IL 60201, USA;\\
$^2$Adler Planetarium, Dept.\ of Astronomy, 1300 S. Lake Shore Drive, Chicago, IL 60605, USA;\\
$^3$Department of Astronomy, University of Wisconsin-Madison, WI 53706, USA;\\
$^4$Space Telescope Science Institute, 3700 San Martin Drive, Baltimore, MD 21218, USA;\\
$^5$Department of Physics, McGill University, McGill Space Institute, 3550 University Street, Montreal, QC H3A 2A7, Canada;\\
$^6$Department of Physics and Astronomy, University of Manitoba, Winnipeg, MB, R3T 2N2, Canada; \\
$^7$Institut f\"ur Astrophysik, Universit\"at G\"ottingen, Friedrich-Hund-Platz 1, 37077 G\"ottingen, Germany; \\
$^8$Department of Astrophysics, American Museum of Natural History, Central Park West and 79th Street, New York, NY 10024;\\
$^9$Department of Physics and Astronomy, McMaster University, Hamilton, ON L8S 4M1, Canada; \\
$^{10}$Visiting	Researcher,	NRC -- Herzberg Astronomy and Astrophysics, National Research Council of Canada, 5071 West Saanich Road, Victoria, British Columbia V9E 2E7, Canada}

\email{$^\dagger$a-geller@northwestern.edu}

\shortauthors{Geller et al.}

\begin{abstract}

Sub-subgiants are stars observed to be redder than normal main-sequence stars and fainter than normal subgiant (and giant) stars in an optical color-magnitude diagram.
The red straggler stars, which lie redward of the red giant branch, may be related and are often grouped together with the sub-subgiants in the literature.
These stars defy our standard theory of single-star evolution, and are important tests for binary evolution and stellar collision models.  
In total, we identify 65 sub-subgiants and red stragglers in 16 open and globular star clusters from the literature; 50 of these, including 43 sub-subgiants,
pass our strict membership selection criteria (though the remaining sources may also be cluster members).
In addition to their unique location on the color-magnitude diagram, we find that at least 58\% (25/43) of sub-subgiants in this sample are X-ray sources 
with typical 0.5-2.5 keV luminosities of order 10$^{30 - 31}$ erg s$^{-1}$.  Their X-ray luminosities and optical-to-X-ray flux ratios are similar to 
those of RS CVn active binaries.
At least 65\% (28/43) of the sub-subgiants in our sample are variables, 21 of which are known to be radial-velocity binaries.  
Typical variability periods are $\lesssim$15 days.
At least 33\% (14/43) of the sub-subgiants are H$\alpha$ emitters.
These observational demographics provide strong evidence that binarity is important for sub-subgiant formation.
Finally, we find that the number of sub-subgiants per unit mass increases toward lower-mass clusters, such that the open clusters in our sample 
have the highest specific frequencies of sub-subgiants.

\end{abstract}

\keywords{open clusters and associations -- globular clusters -- binaries (including multiple): close -- blue stragglers -- stars: evolution -- stars: variables: general}

\section{Introduction}
\label{s:intro}

Sub-subgiant stars are defined empirically as sources that fall redward of the normal main sequence (MS) and are fainter than the subgiant and giant branches 
in an optical color-magnitude diagram (CMD).
Stars in this region were first noted in the open cluster M67 (NGC 2682) by \citet{belloni:98} and then discussed in detail 
in \citet{mathieu:03}.  Around this same time, similar stars were noted in the globular cluster 47 Tuc (NGC 104) by \citet{albrow:01}.  Since then a large number of 
sources that fall into this region in a CMD have been pointed out in various papers in both open and globular clusters, and analyzed to 
varying degrees.  These stars are unexplained by the standard theory of single-star evolution, and despite the growing number of such sources, their origin remains a mystery.

This is the first paper in a series studying the origin of these stars,
formed within both open and globular clusters (and perhaps also in the Galactic field).
In this paper we gather these sources from the many disparate references in the literature, examine their demographics, 
and begin to discuss their possible formation channels. 

First, we address a conflicting naming convention that appears throughout the literature in reference to these stars.  
\citet{belloni:98} introduced the name ``sub-subgiant'' while \citet{albrow:01} used the name ``red straggler''.  
The motivation behind the name sub-subgiants is empirically driven, as these stars are fainter than the normal subgiant branch.  
The term red straggler is also empirically motivated, as these stars are indeed redder than stars of similar luminosity in these clusters.  
The term red straggler is no doubt also influenced
by the more well known ``blue stragglers'' and 
``yellow stragglers/giants''\footnote{Previously, the term red straggler was also used in the literature to refer to stars 
found in between the blue straggler and 
red giant regions on the CMD, sometimes described as evolved blue stragglers \citep[e.g.][]{eggen:83, eggen:88, eggen:89, landsman:97}.
More recently these stars are referred to as ``yellow stragglers'' or, more commonly, ``yellow giants''.}
(both of which are generally brighter than the normal subgiant branch).

In short, depending on the specific reference, a star in the same region of the CMD may be called by either name, and 
therefore there is confusion in the literature about which CMD domain sub-subgiants and red stragglers occupy.

To clarify this nomenclature, we choose to use the term ``sub-subgiant'' (hereafter SSG) to refer to stars that are fainter than the normal 
subgiants and redder than the normal MS stars, shown as the dark-gray regions in Figure~\ref{f:CMDobs}.  
In addition to these SSG stars, there are also stars that are observed to be redder than the normal red giants but brighter than the normal subgiants
(found in the light-gray regions of Figure~\ref{f:CMDobs}). 
We choose to use the term ``red straggler'' (hereafter RS) to refer to these stars, and we encourage the community to adopt this division of the SSG and RS 
nomenclature moving forward. 

We divide the SSG and RS stars in a given photometry band (and isochrone family) by the magnitude at the base of the giant branch, 
as shown in Figure~\ref{f:CMDobs}.  
This definition depends on the position of the isochrone, which might vary if defined by different photometric studies
or isochrone families.  We take cluster parameters from the literature (see Table~\ref{clustab}) and use PARSEC isochrones \citep{bressan:12} here.
For much of our sample, this definition is sufficient to place a star into either the SSG 
or RS category for all of the optical photometry 
we collected from the literature (and provide in Table~\ref{sumtab}).  
However, some stars reside in the SSG region in one color-magnitude combination and in 
the RS region for a different combination (see Figure~\ref{f:CMDobs} and Table~\ref{sumtab}).  We discuss this phenomenon in some detail for specific sources below.  
Briefly, some stars move around dramatically relative to the isochrone with different filter choices, perhaps due to spot activity and
photometric variability (see also, e.g., \citealt{milliman:16}).
Indeed, photometric variability appears to be a defining characteristic of the SSG and RS stars.
For the SSG analysis presented here, we include all stars that fall in the SSG region in at least one optical color-magnitude 
combination.  This is the most inclusive definition of SSGs, which may be important given the photometric variability.
We will also account for potential field star contamination within this SSG sample (Section~\ref{s:contam}) prior to performing our demographic analyses.

In this paper, we gather the observations from the SSG and RS stars in each cluster in Section~\ref{s:obs} and provide a summary 
of the observations of these stars in Table~\ref{sumtab}.  In Sections~\ref{s:bias} and~\ref{s:contam} we discuss possible observational
biases in our sample and the potential for field-star contamination, respectively.  
In Section~\ref{s:agg} we investigate the aggregate characteristics of the sample.  We close with a brief discussion and conclusions in 
Section~\ref{s:discuss}.  Subsequent papers in this series will investigate in detail a set of theoretical formation channels for 
SSGs (some of which predict that SSG and RS stars may be related through formation and/or evolution), 
and evaluate the formation rates of each channel and their abilities to create SSGs with the demographics identified here.

\section{Observed Sub-subgiants}
\label{s:obs}

In Table~\ref{sumtab} we compile all of the SSG and RS stars that have been identified in the literature in both open and globular clusters.   We provide the 
cluster's NGC name, another common name where appropriate, the SSG/RS ID  (see the relevant paragraph below for the specific references for IDs and other values), 
the RA and Dec, proper-motion and radial-velocity memberships statuses (P$_\text{PM}$ and P$_\text{RV}$, where available), our estimate of the probability that this 
star is a field star (P$_\text{field}$, see Section~\ref{s:contam}), the radial distance from the cluster center in 
units of core radii ($r_\text{c}$), the available observed $UBVRIJHK$ photometry\footnote{All optical photometry are given in 
Johnson-Cousins filters (and have not been de-reddened).  Where necessary, \textit{HST} magnitudes are converted to 
ground-based Johnson-Cousins following \citet{holtzman:95} and \citet{sirianni:05}.  All $JHK$ infrared photometry come from 2MASS \citep{skrutskie:06}. },
an available X-ray luminosity ($L_X$) and the band of the X-ray observation, the radial-velocity orbital period (``Per$_\text{RV}$'') and photometric period (``Per$_\text{phot}$''),
where available (we mark variables whose periods are yet to be determined as ``var''), and finally the number of color-magnitude 
combinations given in this table (from the literature) that place the given star in the SSG or RS regions or neither (``SSG/RS/N''). 

In Table~\ref{sumtab} and throughout this paper, we 
identify sources with velocities that are $>3\sigma$ from the cluster mean
(if that is the only available membership measurement) or with formal membership probabilities $<$50\% as non-members.
We exclude definite non-members from Table~\ref{sumtab}.
In some cases, authors identify stars as non-members at $<3\sigma$ from the mean (e.g., at $>2\sigma$ or $>2.5\sigma$).  
We choose to include such stars with questionable membership within Table~\ref{sumtab}, and indicate this uncertain membership with ``?'', 
but we do not include these in our subsequent analysis.  We discuss additional membership indicators in Section~\ref{s:contam}.

\begin{figure*}[!t]
\begin{tabular}{cccc}
\includegraphics[width=0.23\linewidth]{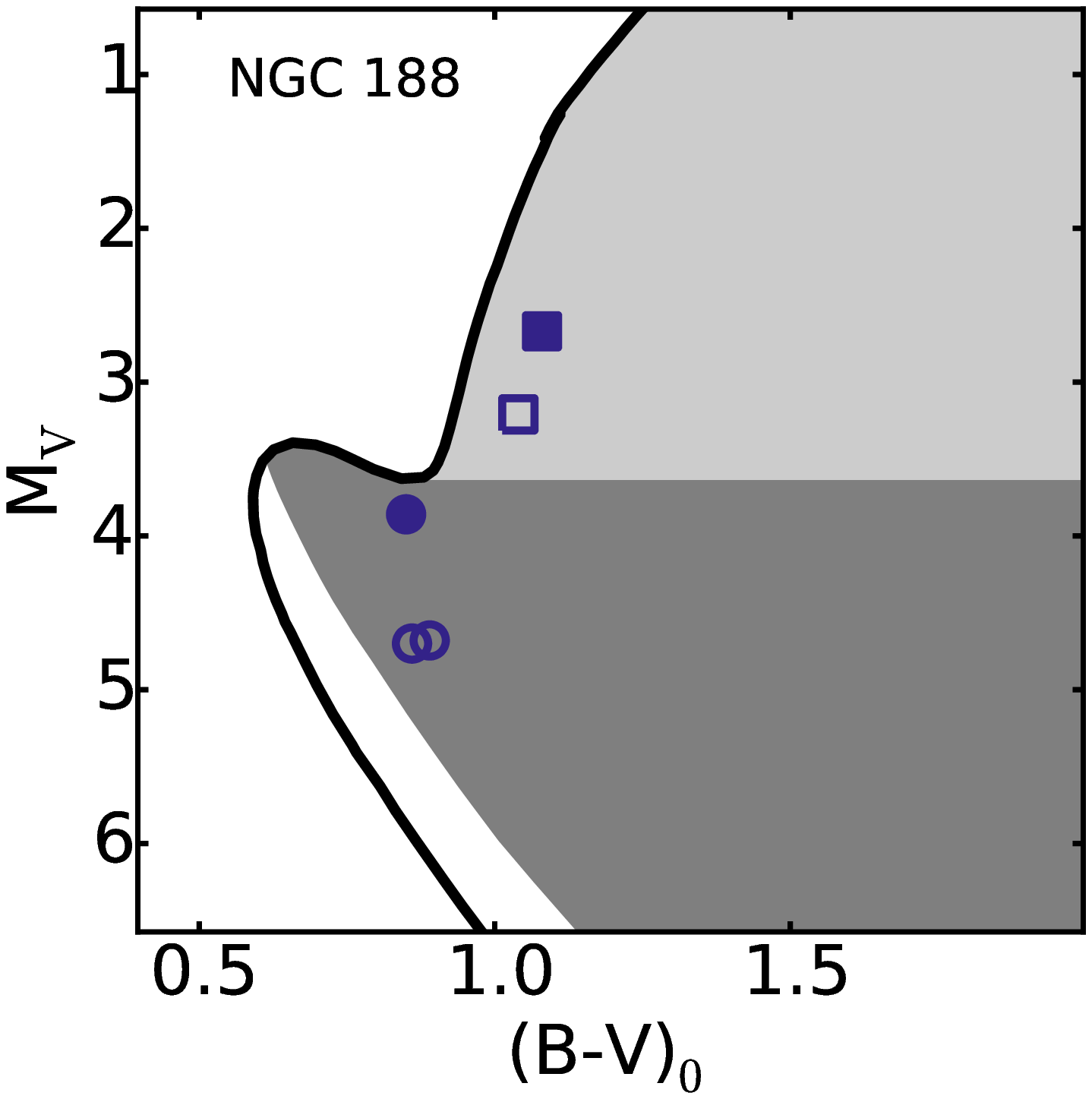} & \includegraphics[width=0.23\linewidth]{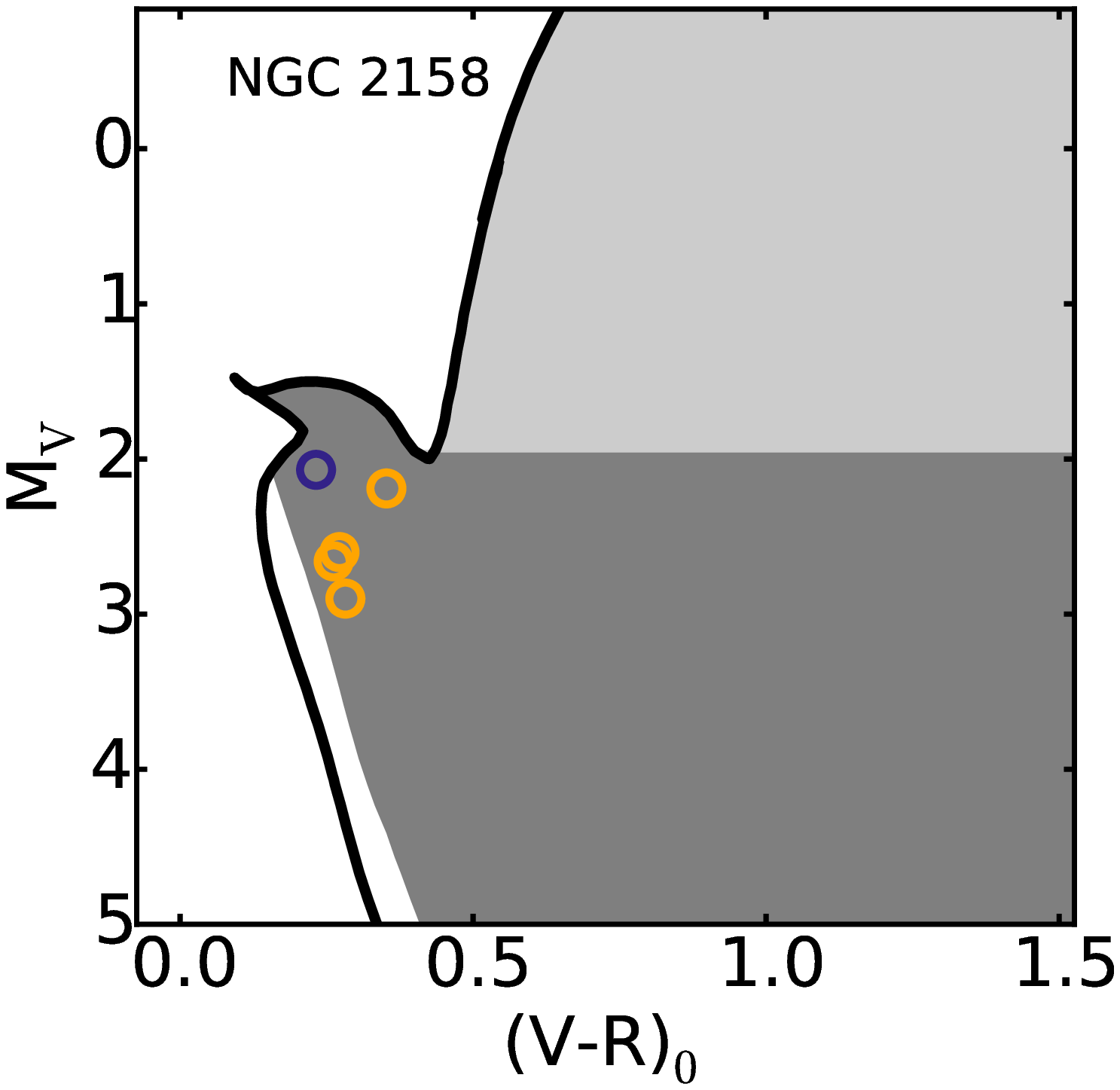} & \includegraphics[width=0.23\linewidth]{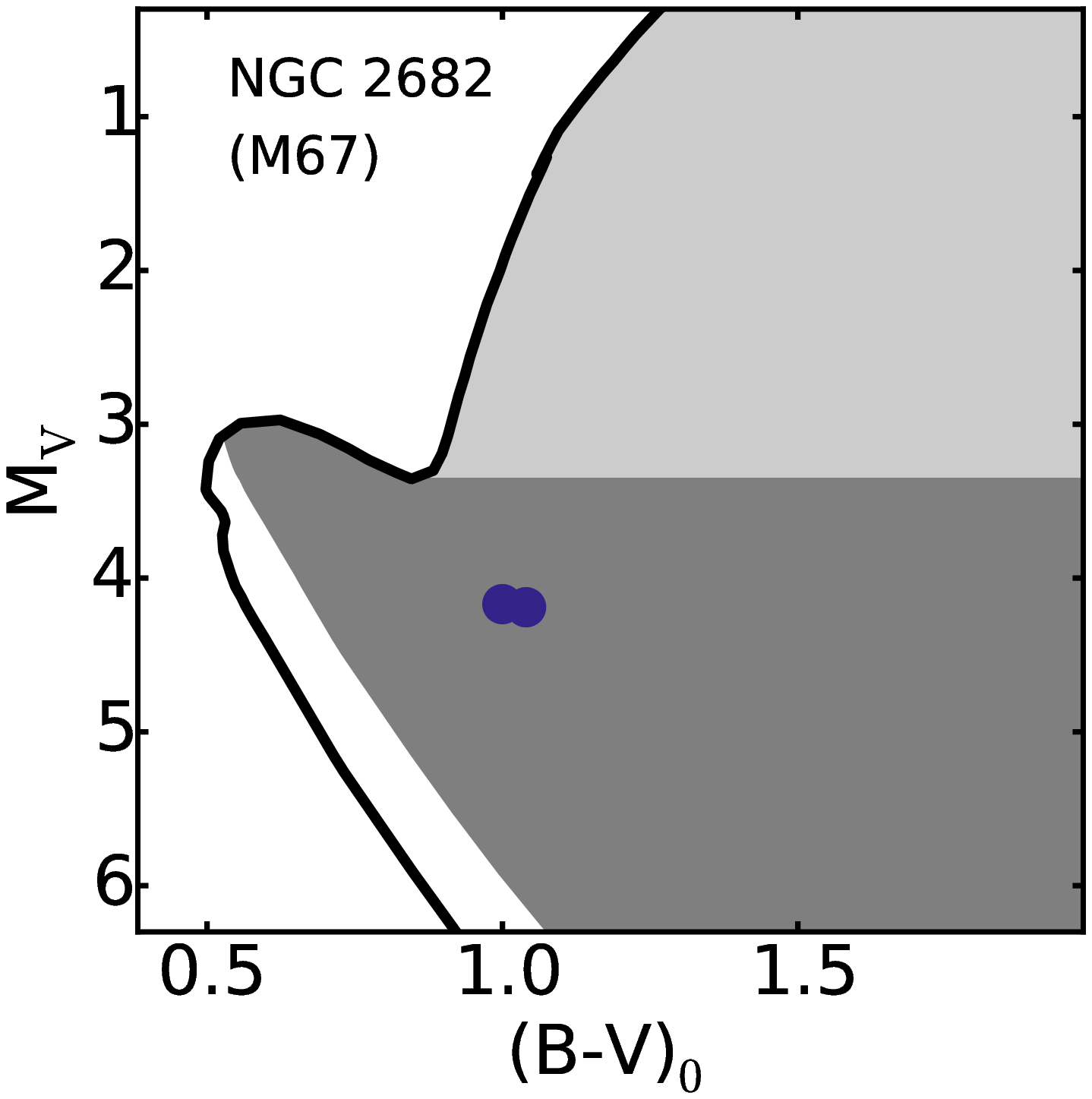} & \includegraphics[width=0.23\linewidth]{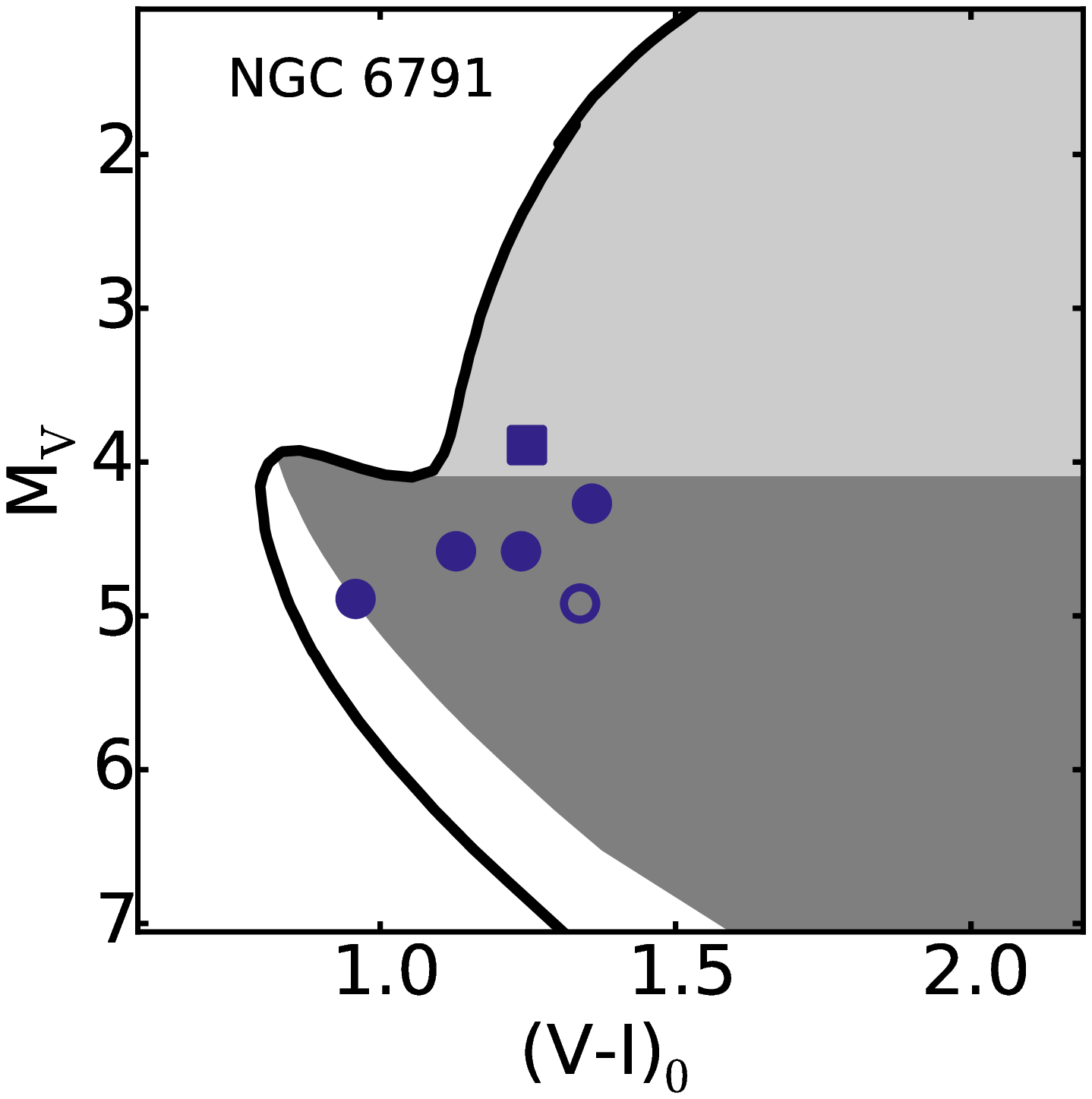} \\
\includegraphics[width=0.23\linewidth]{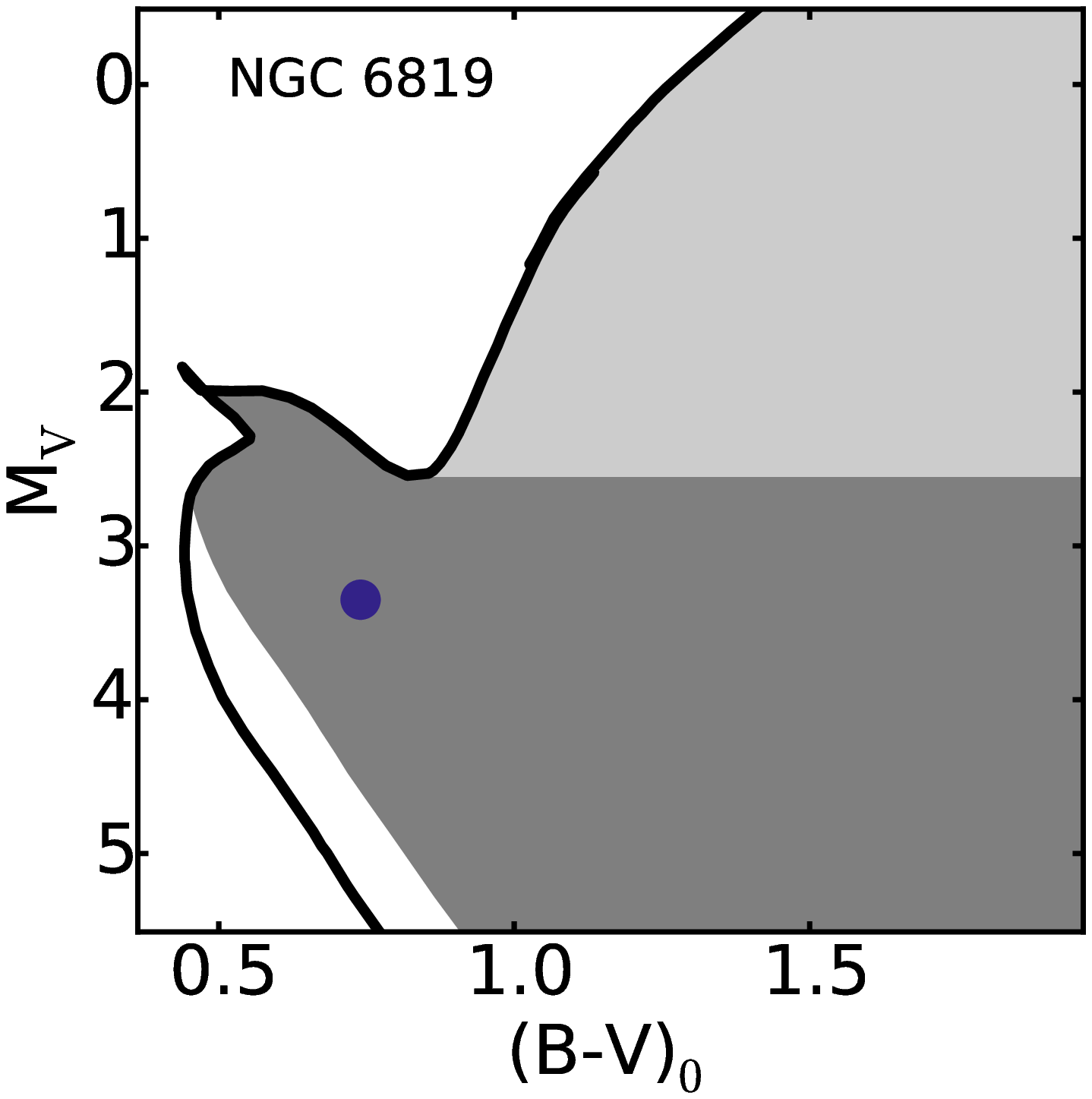} & \includegraphics[width=0.23\linewidth]{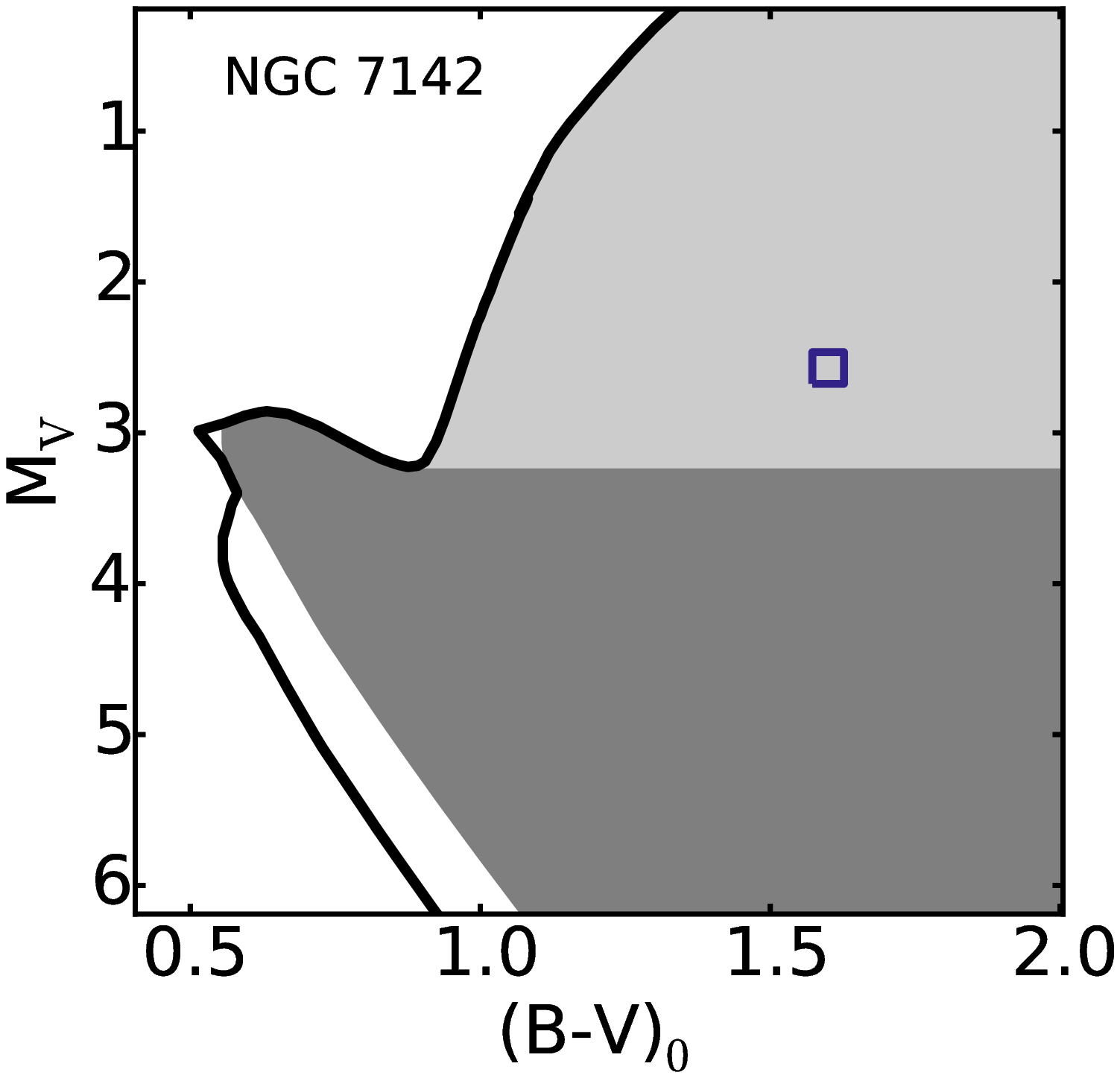} & \includegraphics[width=0.23\linewidth]{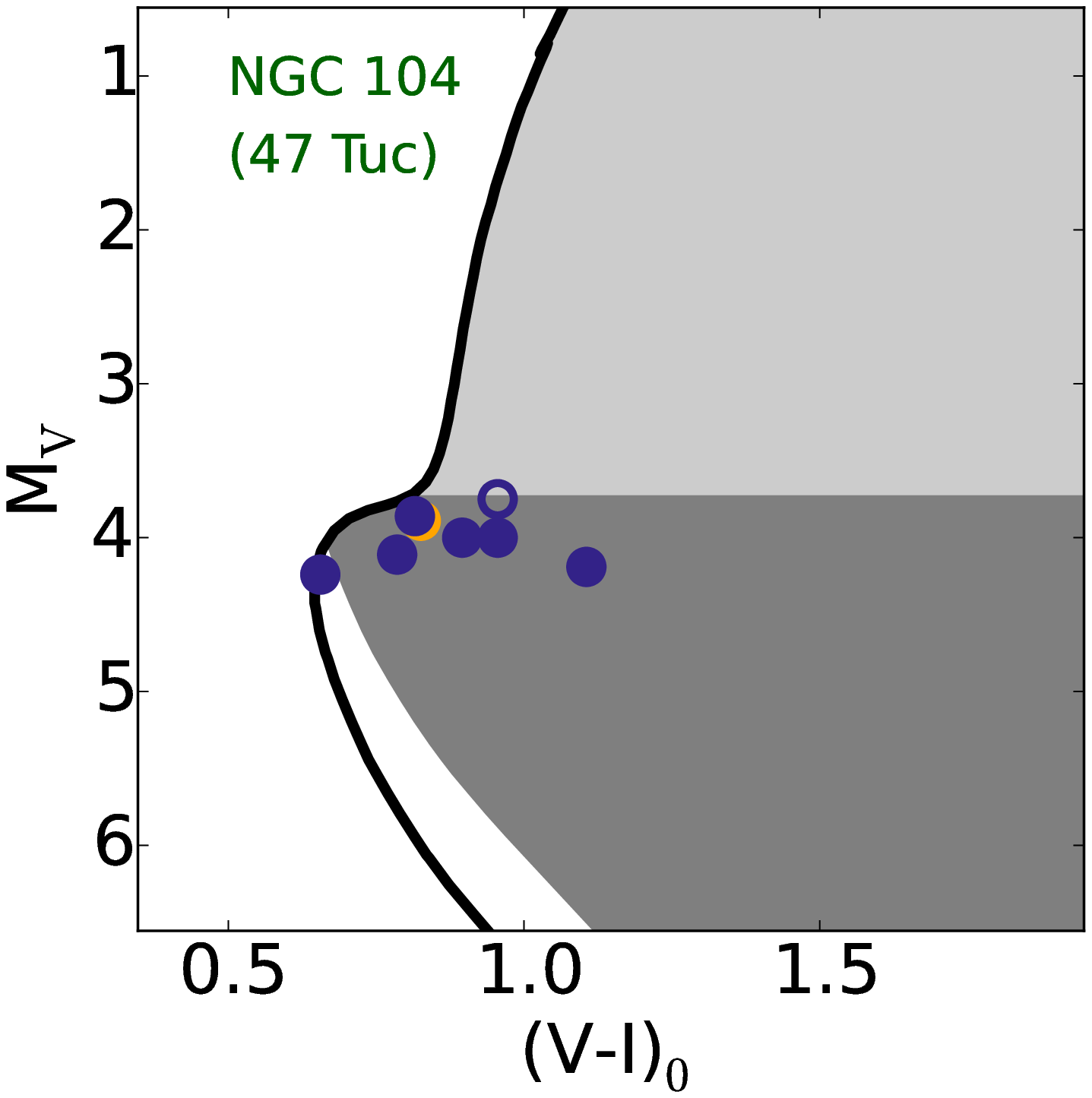} & \includegraphics[width=0.23\linewidth]{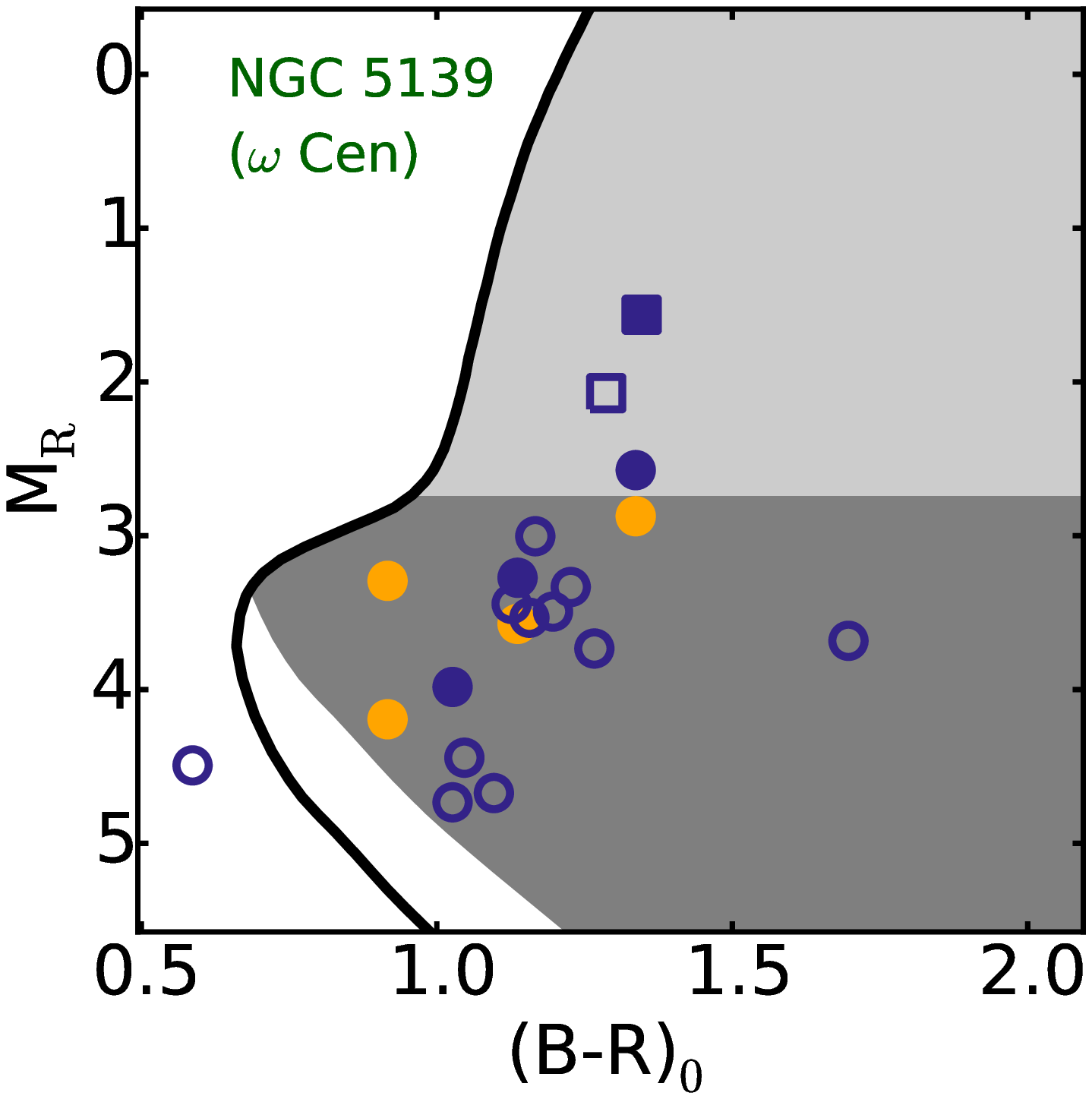} \\
\includegraphics[width=0.23\linewidth]{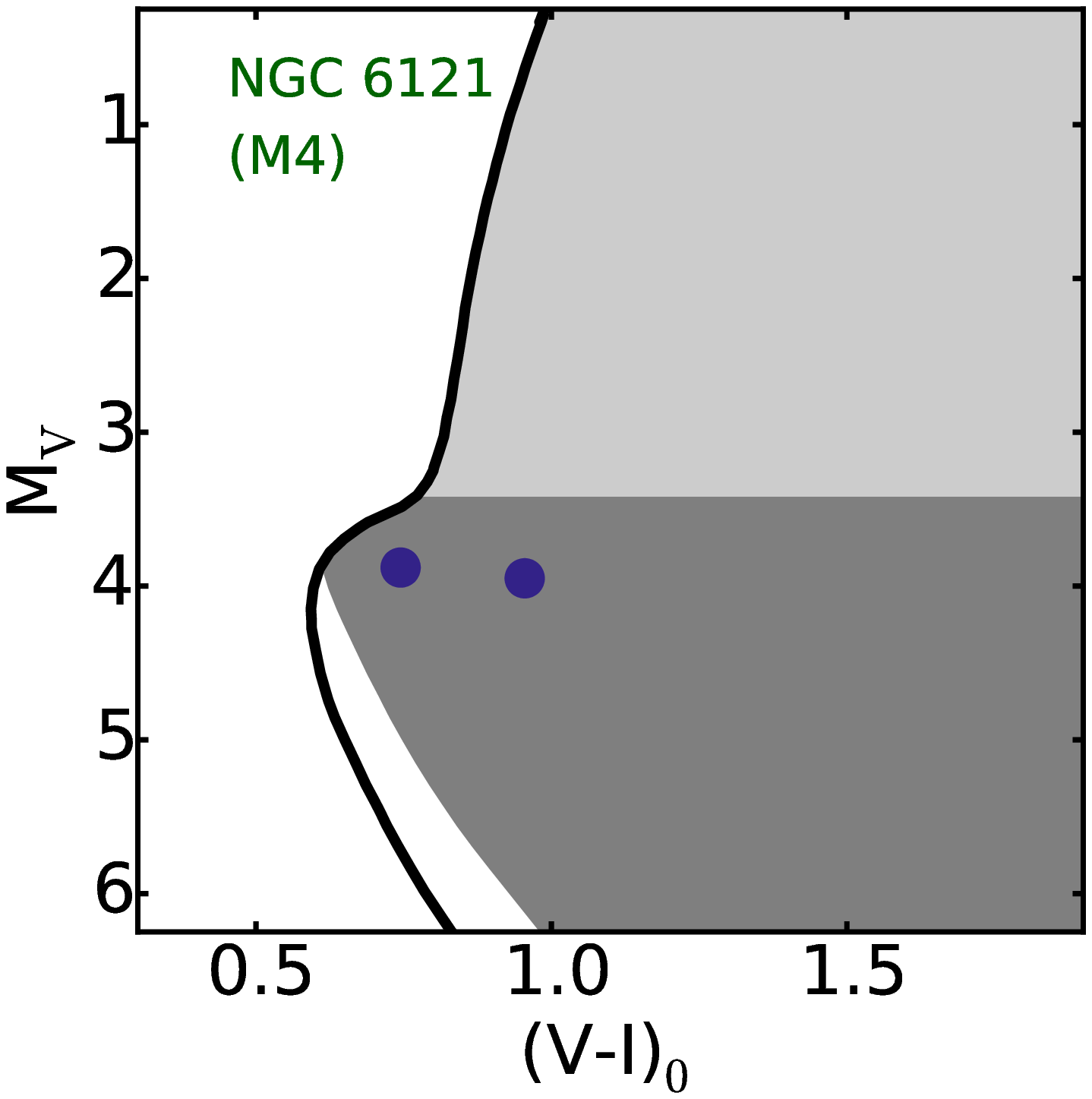} & \includegraphics[width=0.23\linewidth]{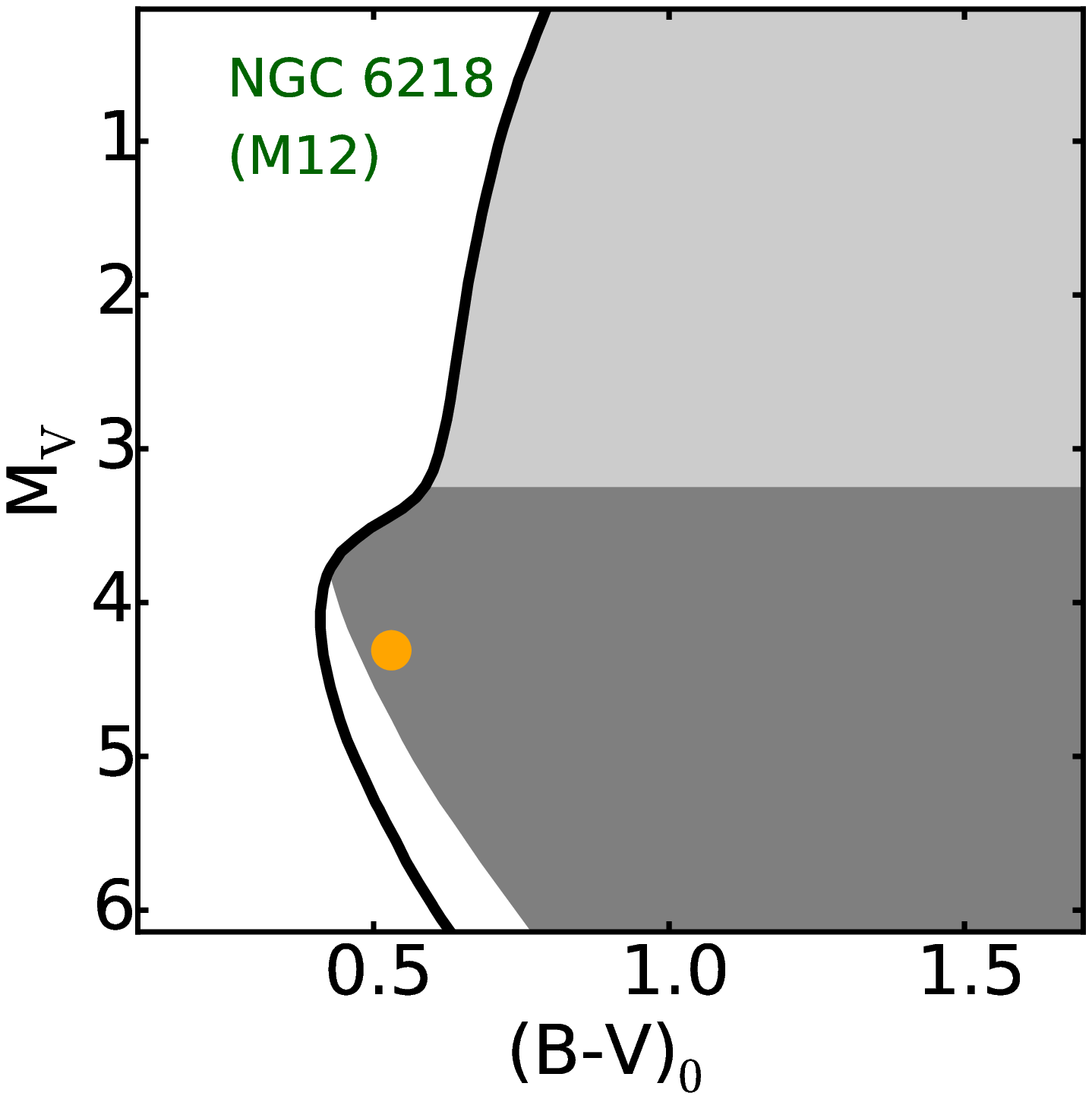} & \includegraphics[width=0.23\linewidth]{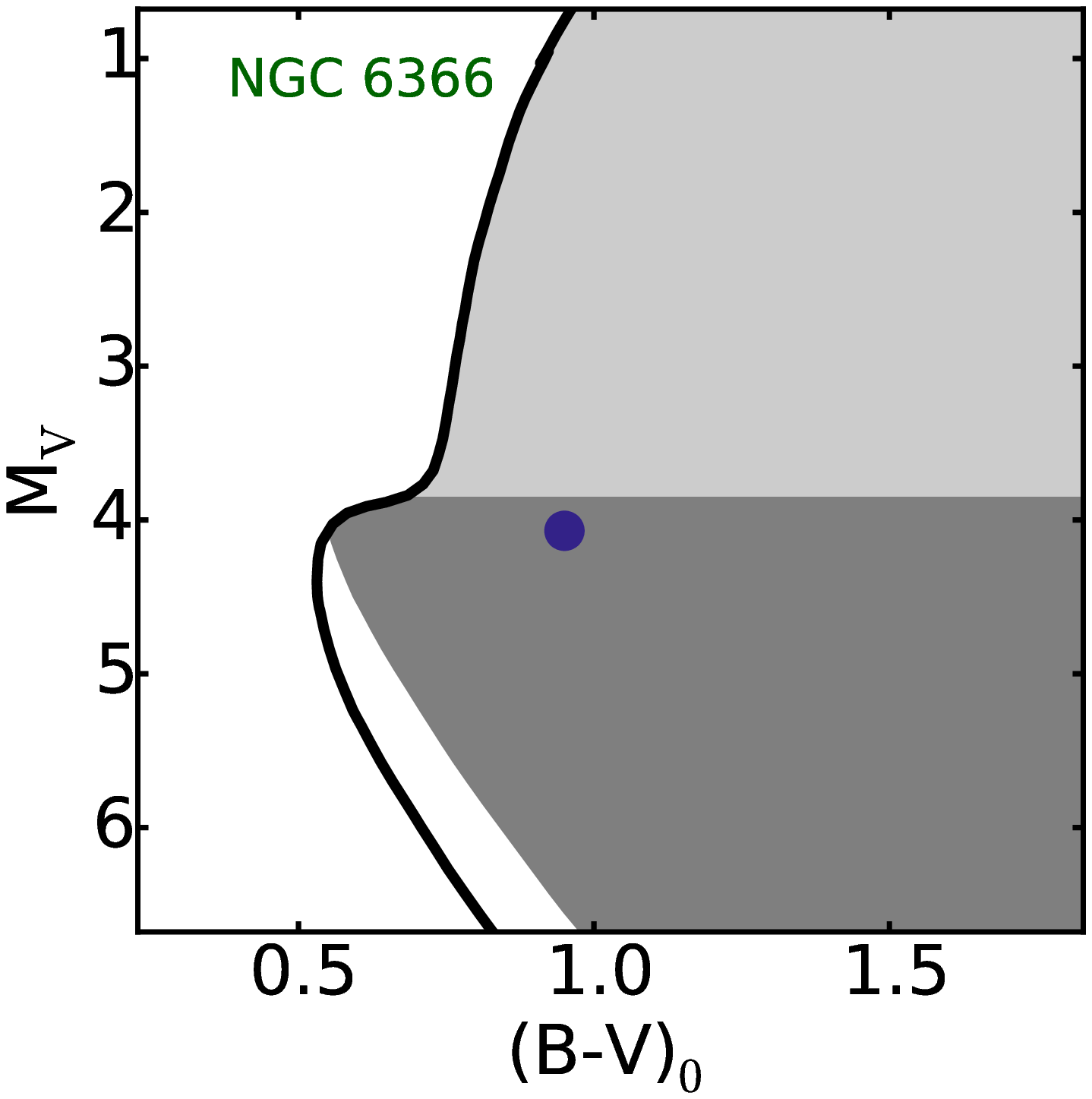} & \includegraphics[width=0.23\linewidth]{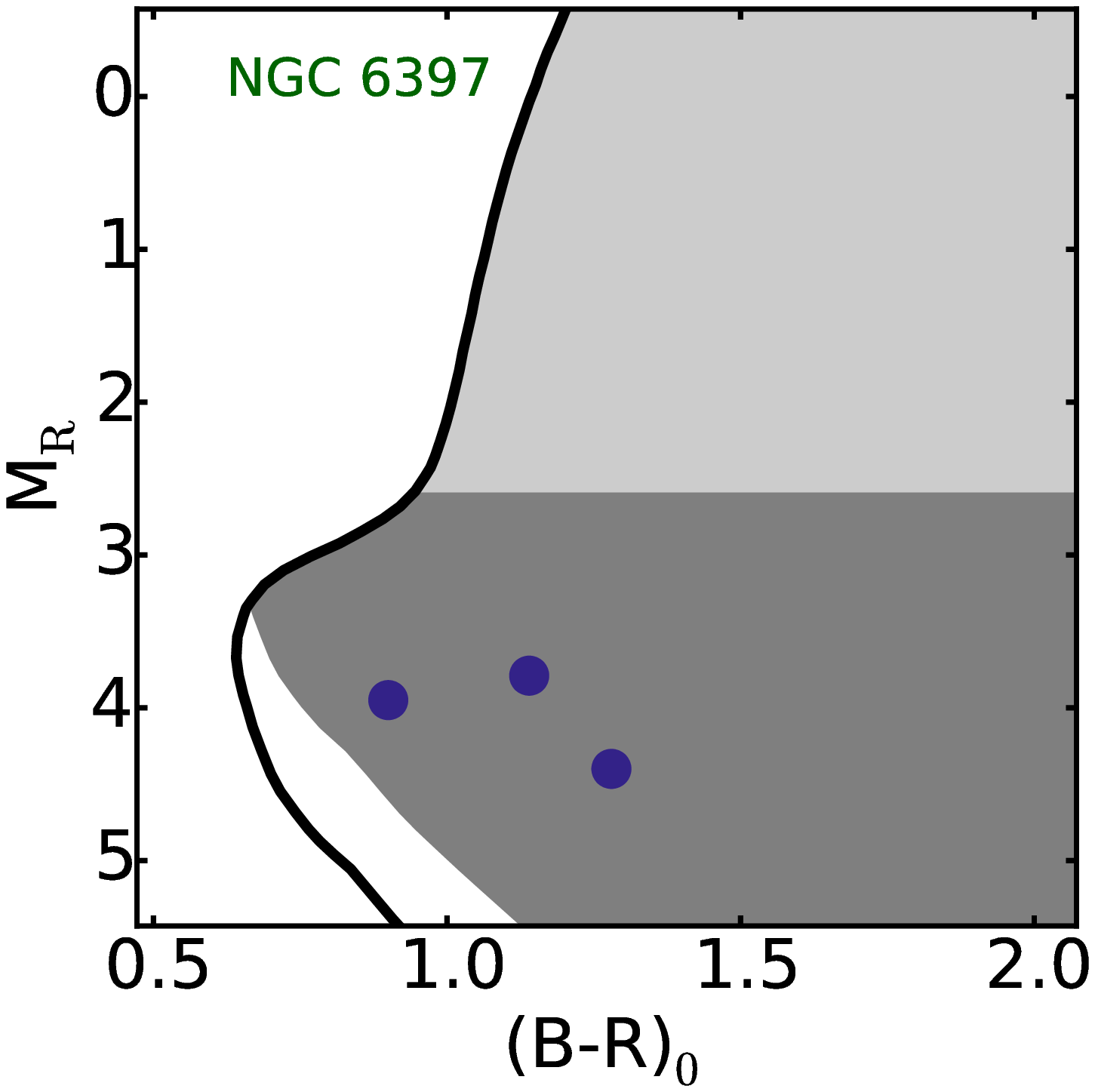} \\
\includegraphics[width=0.23\linewidth]{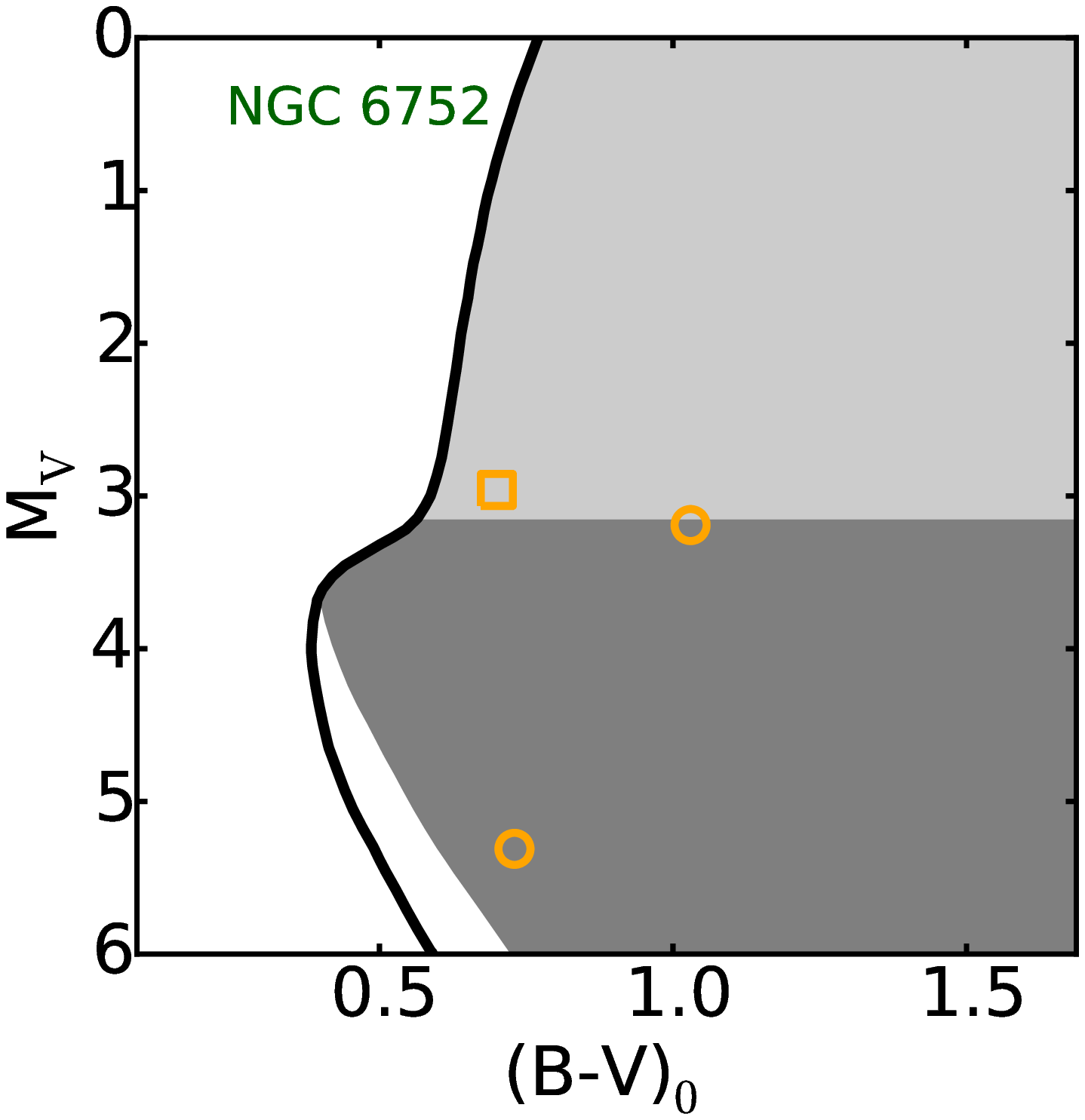} & \includegraphics[width=0.23\linewidth]{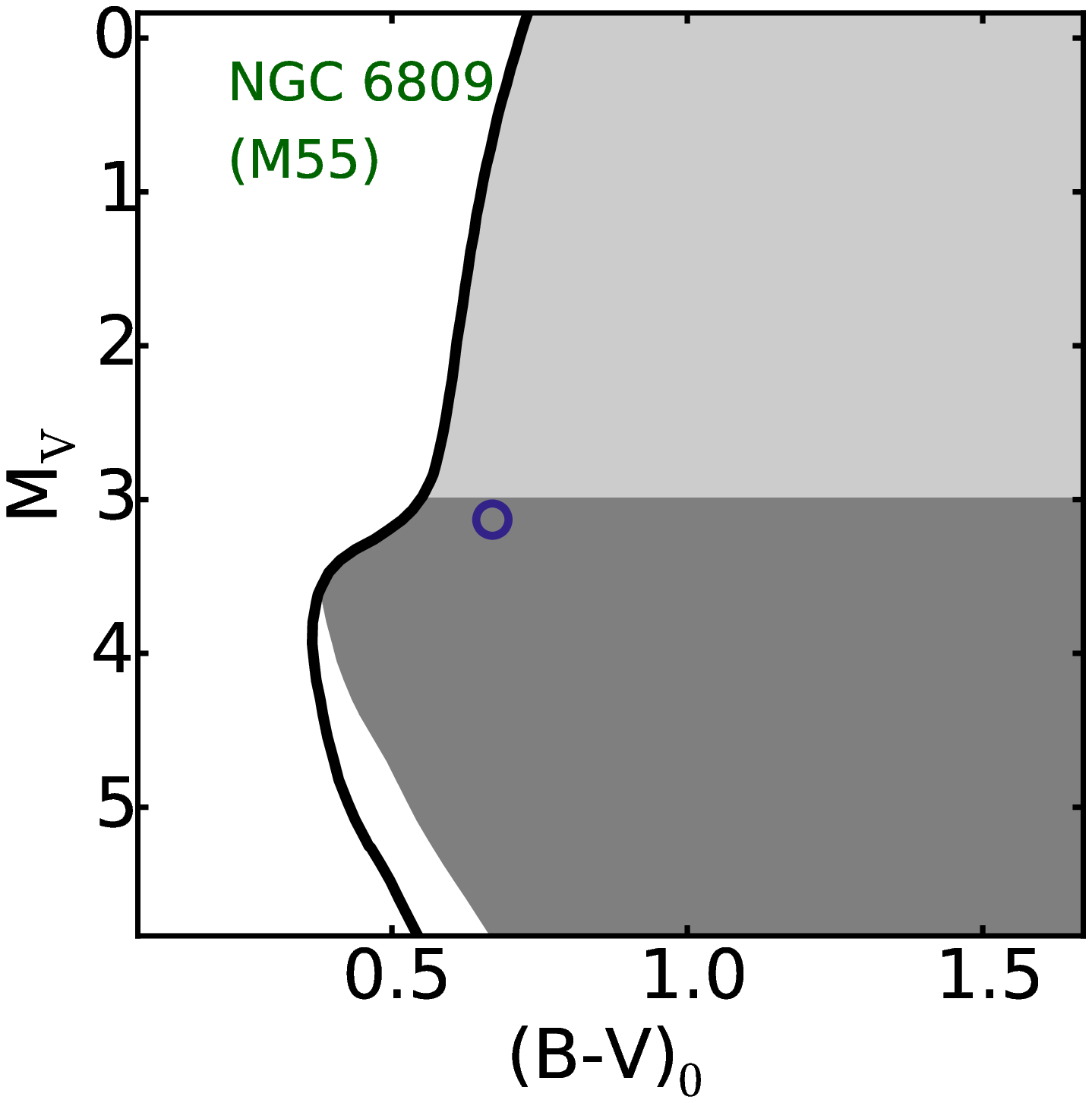} & \includegraphics[width=0.23\linewidth]{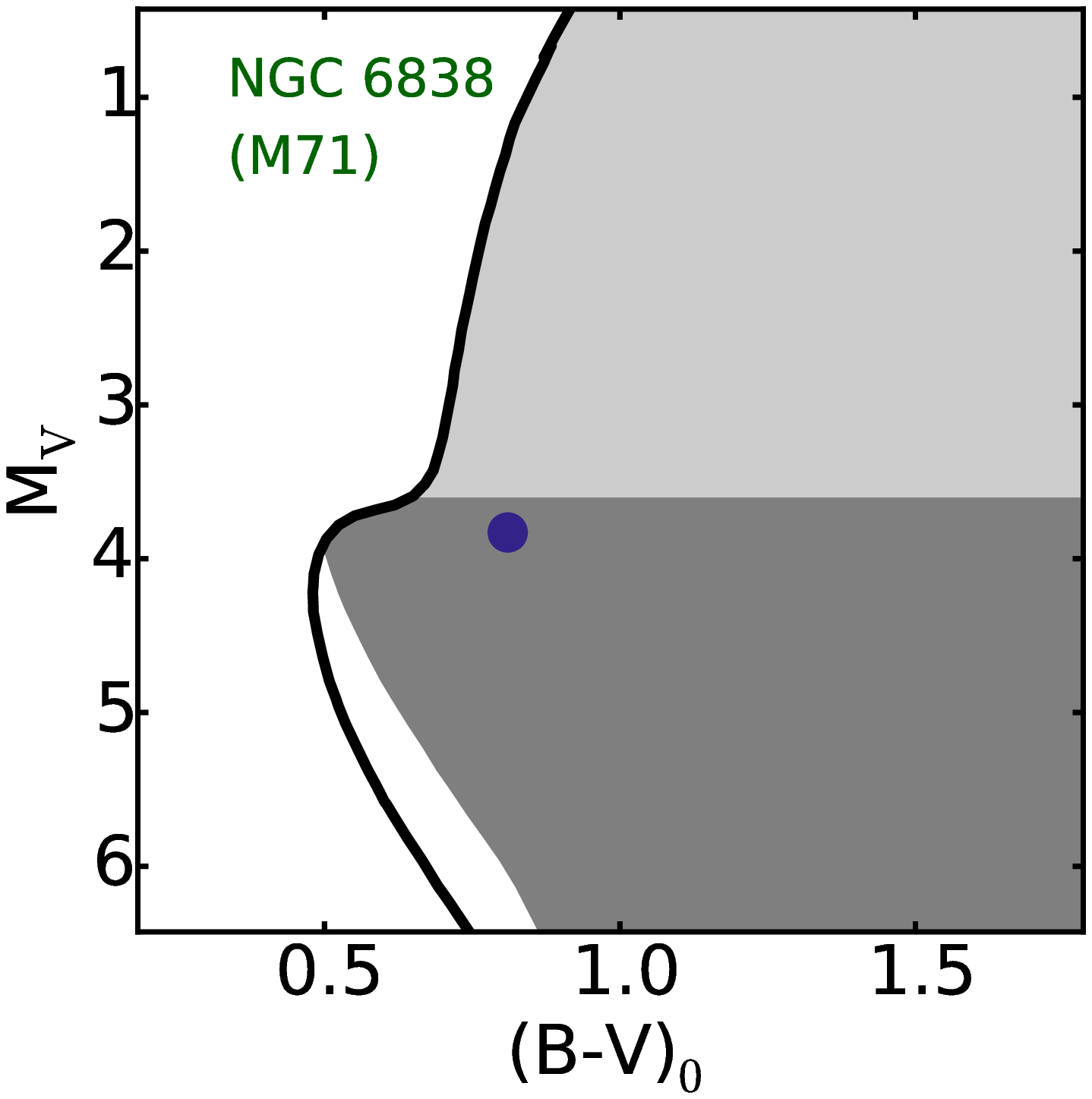} & \includegraphics[width=0.23\linewidth]{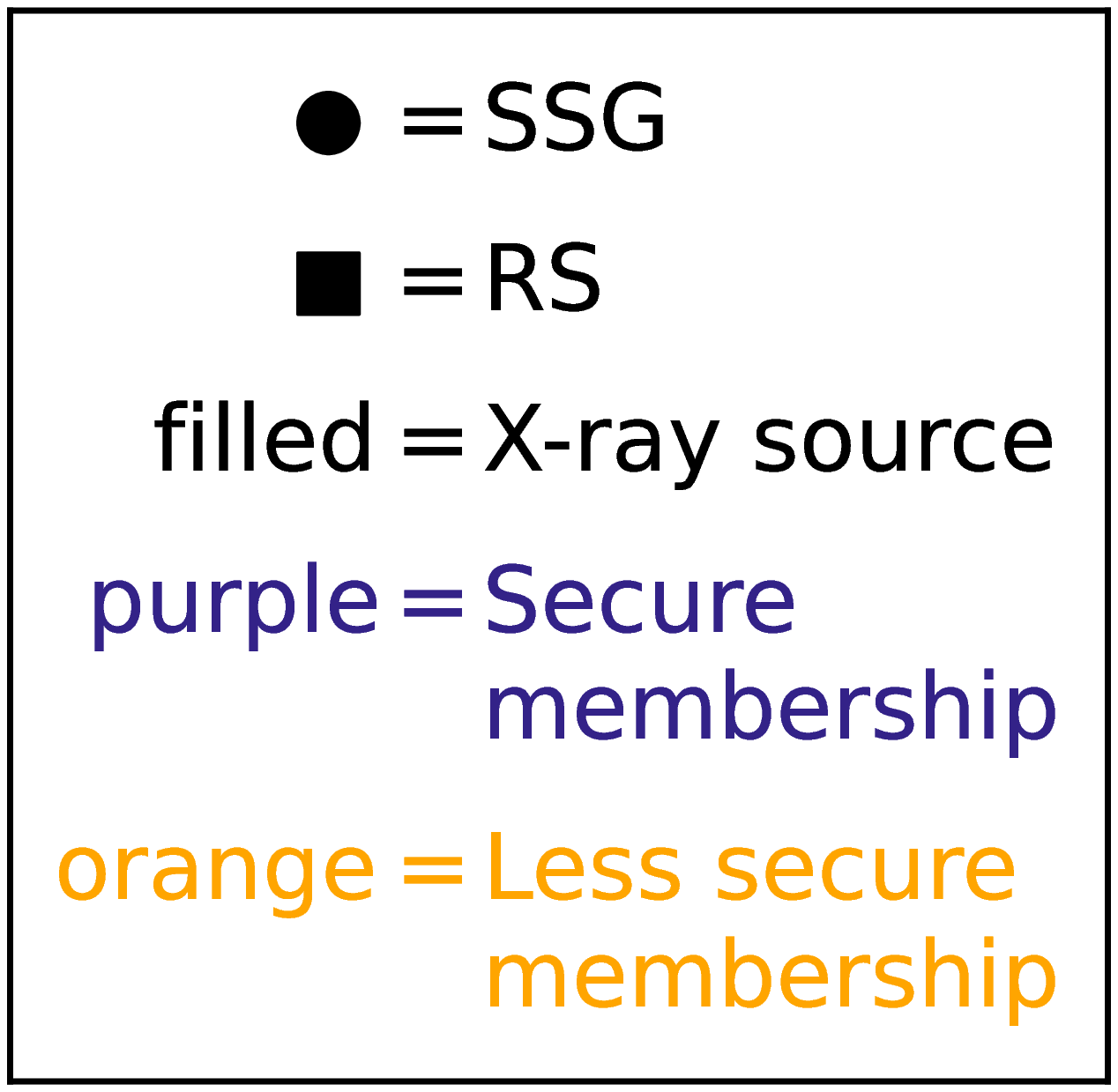} \\
\end{tabular}
\caption{
Color-magnitude diagrams for the SSG and RS stars in each respective cluster.  
Open clusters are plotted first (with names colored black) followed by the globular clusters (with names colored dark green).
All magnitudes are given in Table~\ref{sumtab}, and, where necessary, we choose the color-magnitude combination for a given cluster that allows us to 
plot the largest number of sources.  For reference we also plot with black lines PARSEC isochrones \citep{bressan:12}, using the ages, 
distance moduli and reddening values from Table~\ref{clustab}.  The SSG region is shown 
in the dark-gray filled area, while the RS region is shown in the light-gray filled area (both as defined in Section~\ref{s:intro}, and 
note that we exclude the normal binary locus from these regions.)
We plot sources that appear in the SSG region in at least one color-magnitude combination with circles, and the RS stars in squares.
Filled symbols show known X-ray sources, and open symbols show those without detected X-ray emission. 
Highly likely cluster members are plotted with purple symbols; those with less secure membership are plotted in orange 
(see Section~\ref{s:agg}). 
Note that NGC 6652 is not shown here, as we only have an estimate of the source's $V$ magnitude (due to high-frequency variability).
\label{f:CMDobs}
}
\end{figure*}

Observations of the SSG stars reveal an intriguing mixture of properties.  Broadly, SSGs 
\begin{enumerate}
\item are redder than normal MS stars, but fainter than normal giants in an optical color-magnitude diagram,
\item have X-ray luminosities, $L_X$, of order $10^{30} - 10^{31}$~erg~s$^{-1}$ in both open and globular clusters,
\item are often H$\alpha$ emitters (where measurements have been made),
\item exhibit photometric variability with periods $\lesssim 15$ days (where available), and
\item where possible are mostly identified as radial-velocity binaries.
\end{enumerate}
We return to these aggregate properties in Section~\ref{s:agg}.
First we briefly discuss the observations from each of the individual star clusters listed in Table~\ref{sumtab}.
CMDs for each of these clusters, showing the SSG and RS stars along with an isochrone (for reference) and the SSG/RS regions are plotted in Figure~\ref{f:CMDobs}, 
and relevant cluster parameters are shown in Table~\ref{clustab}.

\begin{deluxetable*}{lccccccc}
\tabletypesize{\tiny}
\tablecaption{Cluster Parameters \label{clustab}}
\tablehead{\colhead{Cluster} & \colhead{age} & \colhead{$M_\text{cl}$} & \colhead{$(m-M)_V$} & \colhead{$E(B-V)$} & \colhead{[Fe/H]} & \colhead{$n_{\rm H}$} & \colhead{$(r/r_c)_\text{max}$} \\
\colhead{} & \colhead{[Gyr]} & \colhead{[\Msolar]} & \colhead{} & \colhead{} & \colhead{} & \colhead{[10$^{20}$cm$^{-2}$]} & \colhead{} }
\tablewidth{0pt}
\startdata
&&&&\\
\multicolumn{5}{l}{\textbf{Open Clusters}} \\
&&&&\\
NGC 188    &               6.2   &  1500 & 11.44   &   0.09   &             0.0  &  6.6 & 7.1 \\
NGC 2158   &                 2   & 15000 & 14.51   &   0.55   &            -0.6  & 41.9 & 7.9 \\
NGC 2682   &                 4   &  2100 &   9.6   &   0.01   &             0.0  &  3.3 & 7.2 \\
NGC 6791   &                 8   &  4600 & 13.38   &    0.1   &             0.4  & 10.7 & 10.6 \\
NGC 6819   &               2.4   &  2600 &  12.3   &    0.1   &             0.0  & 19.3 & 12.5 \\
NGC 7142   &               3.6   &   500 & 12.86   &   0.29   &             0.1  & 32.6 & 2.2 \\
&&&&\\
\multicolumn{5}{l}{\textbf{Globular Clusters}} \\
&&&&\\
NGC 104    &            13.1   &  1.0$\times$10$^6$ & 13.37   &   0.04   &           -0.72  &  5.4 & 3.6 \\
NGC 5139   &            11.5   &  2.2$\times$10$^6$ & 13.94   &   0.12   &           -1.53  &  8.8 & 5.1 \\
NGC 6121   &            12.5   &  1.3$\times$10$^5$ & 12.82   &   0.35   &           -1.16  & 14.2 & 3.3 \\
NGC 6218   &            12.7   &  1.4$\times$10$^5$ & 14.01   &   0.19   &           -1.37  &  7.7 & 8.9 \\
NGC 6366   &            13.3   &  4.8$\times$10$^5$ & 14.94   &   0.71   &           -0.59  & 14.0 & 5.5 \\
NGC 6397   &            12.7   &  7.7$\times$10$^4$ & 12.37   &   0.18   &           -2.02  & 11.4 & 50.8 \\
NGC 6652   &            12.9   &  7.9$\times$10$^4$ & 15.28   &   0.09   &           -0.81  &  9.2 & 40.0 \\
NGC 6752   &            11.8   &  2.1$\times$10$^5$ & 13.13   &   0.04   &           -1.54  &  4.9 & 41.1 \\
NGC 6809   &            12.3   &  1.8$\times$10$^5$ & 13.89   &   0.08   &           -1.94  &  9.4 & 6.5 \\
NGC 6838   &            12.0   &  3.0$\times$10$^4$ & 13.84   &   0.25   &           -0.78  & 21.2 & 7.8 \\
\enddata
\tablenotetext{}{Note: References for the values in this table are as follows. For the open clusters: 
NGC 188: \citet{sarajedini:99}, \citet{meibom:09}  and \citet{chumak:10}.
NGC 2158: \citet{carraro:02}.
NGC 2682: \citet[][and references therein]{geller:15}.
NGC 6791: \citet{Stetson2003}, \citet{carney:05} and \citet{tofflemire:14}
NGC 6819: \citet{kalirai:01} and \citet[][and references therein]{hole:09}.
NGC 7142: \citet[][and references therein]{sandquist:13} and \citet{straizys:14}.
For the globular clusters we take the age from \citet[][using the ``G00$_{\rm CG}$'' values and normalized using the age of 47 Tuc from \citealt{thompson:10}]{marin:09}, $(m-M)_V$, $E(B-V)$, [Fe/H] and $M_\text{cl}$ (calculated assuming a mass-to-light ratio of 2) from \citet{harris:96,harris:10}. For NGC 6838 we take the age, $(m-M)_V$, $E(B-V)$ from \citet{dicecco:15}.  All $n_{\rm H}$ values are derived from NASA's HEASARC nH tool (https://heasarc.gsfc.nasa.gov/cgi-bin/Tools/w3nh/w3nh.pl), which uses \citet{dickey:90} and \citet{kalberla:05}.
Finally, note that NGC 6397 and NGC 6752 are core-collapsed clusters; the radial limits of these surveys in 
units of half-mass radii are 0.9 and 3.7, respectively.
}

\end{deluxetable*}

\vspace{1em}
\subsection{Open Cluster Observations}

\paragraph{NGC 188 } This old ($\sim$6-7 Gyr) open cluster has three SSGs and two RSs \citep[IDs from][]{geller:08}.
The optical and IR photometry for these stars in Table~\ref{sumtab} come from \citet{stetson:04} and 2MASS, respectively.
All of these stars have proper motions \citep{platais:03} consistent with cluster membership, and, where measurements are possible,
these stars are also radial-velocity members \citep{geller:08}. 
All but one of these stars (1141, a RS) have radial-velocity variations indicative of a binary companion.
Two of these five sources (one SSG and one RS) are detected as X-ray emitters \citep{belloni:98, gondoin:05}.
\citet{gondoin:05} suggests that the X-rays from 4289 (a SSG) are due to rapid rotation and resulting chromospheric activity.  This source lies very close to 
the base of the giant branch in 
a $V$ vs.\ $B-V$ CMD, but is farther removed from the ``normal'' stars (and isochrones) in CMDs using other photometry combinations,
particularly in combination with $I$ band (given the photometry from \citealt{stetson:04}).  Furthermore, 4289 is a member of a binary with a period of 
11.49 days \citep{geller:09}, which is consistent with the hypothesis that the X-rays result from rapid rotation in the synchronized primary member of the binary.
1141 is a RS and also an X-ray source \citep[][X5]{belloni:98}, though the authors of that study do not venture to guess the source of the emission.  
We note that \citet{geller:08} incorrectly matched 3118 (a RS) to a \citet{belloni:98} X-ray source; to our knowledge 3118 does 
not have detected X-ray emission.  3118 is a double-lined spectroscopic binary (SB2) with a period of 11.9 days and a mass ratio of 0.8 \citep{geller:09}.
Interestingly, SSG 4989 is identified as a W UMa photometric variable, V5, by \citet{zhang:02}, which in general are thought to be contact binaries 
containing two MS stars \citep{robertson:77}. 

\paragraph{NGC 2158 } As part of their search for transiting planets in this intermediate age ($\sim$2 Gyr) open cluster, \citet{mochejska:04,mochejska:06} 
identified five photometric variables in the cluster's SSG region.  
IDs and optical photometry in Table~\ref{sumtab} for these stars are from \citet{mochejska:04,mochejska:06}.
One of these sources (V90) is in the catalog of \citet{dias:14} with a 94\% proper-motion 
membership. The remaining four have proper motions from \citet{kharchenko:97}; all are $>$50\% proper-motion members when considering
the cluster ``corona'' stellar distribution, though all but one of these fall to $<$50\% when considering the ``core'' distribution.
(\citealt{kharchenko:97} describe the cluster as a combination of two distributions, with the ``corona'' having a characteristic radius of twice that of the ``core''.)  
The photometric periods for these sources range from $<$1 day to $\sim$13 days.  To our knowledge, there is no published X-ray survey of NGC 2158.  

\paragraph{NGC 2682 (M67) } \citet{mathieu:03} performed an extensive observational analysis of the two SSGs in the old ($\sim$4 Gyr) open cluster M67, 
and we refer the reader to this paper for more information. 
In short, these stars were first noted by \citet{belloni:98}, and both sources are high-probability cluster members from 
proper motions \citep{girard:89}
and radial velocities \citep{mathieu:03,geller:15}.  Both are radial-velocity binaries (see Table~\ref{bintab}); 
the shorter-period source, S1113 \citep[IDs for both sources from][]{sanders:77}, 
is an SB2 with a companion that is likely a 0.9 \Msolar\ MS star, while S1063, the longer-period source, is a single-lined spectroscopic binary (SB1). 
Both of these sources are photometric variables \citep{vandenberg:02} and X-ray sources \citep{belloni:98,vandenberg:04}.  
(The optical and IR photometry in Table~\ref{sumtab} for these stars is from \citep{montgomery:93} and 2MASS, respectively.)
As noted in \citet{mathieu:03}, both stars show strong Ca II H and K emission, indicative of 
chromospheric activity, and both also show H$\alpha$ emission \citep{pasquini:98,vandenberg:99}. 
Finally, \citet{mathieu:03} note that they could not find a self-consistent solution for the stars of S1113 that accounts for all of the observations.

\paragraph{NGC 6791 } \citet{Platais2011} identify five stars in the SSG region, and \citet{vandenBerg2013}
identify two additional SSG/RS stars (6371 and 7011) as optical counterparts to X-ray sources in this old ($\sim$8 Gyr), 
metal rich ([Fe/H]= 0.4) open cluster.
The IDs and optical photometry for these sources are from \citet{Stetson2003}, and the IR photometry is from 2MASS.
All of these sources have proper-motion membership probabilities of $P_{PM} \geq 96\%$ \citep{Platais2011}.
Four of the five \citet{Platais2011} candidates (83, 746, 3626, 15561) were confirmed to be radial-velocity members by \citet{milliman:16}.  
The authors are currently collecting radial-velocity measurements for the other two SSG candidates through the 
WIYN Open Cluster Study \citep[WOCS;][]{mathieu:00}.
Though not published in \citet{milliman:16}, these WOCS radial velocities indicate that
6371 and 7011 are both short-period SB2 binaries with radial velocities spanning the cluster distribution;
we have not yet been able to derive orbital solutions for these stars, so we cannot yet provide conclusive radial-velocity membership probabilities.
The five \citet{Platais2011} sources are also short-period radial-velocity binaries.
\citet{milliman:16} published orbital solutions for 746, 3626 and 15561 (see Table~\ref{bintab}).
Five of these NGC 6791 sources (four SSGs and one RS) are short-period photometric variables, X-ray sources and H$\alpha$ emitters 
\citep{deMarchi2007, Mochejska2002, Kaluzny2003, Bruntt2003, Mochejska2005, vandenBerg2013, milliman:16}.
For the variable SSG and RS sources in NGC 6791, photometric variability occurs on periods similar to the radial-velocity orbital period and 
has been attributed to spot modulation \citep{vandenBerg2013}. All signs point to these stars being RS CVn-type
binaries with chromospheric activity. On the other hand, star 83
shows no signs of a binary companion, photometric variability, H$\alpha$ 
emission, or X-ray emission, so it appears qualitatively different than the other SSGs in the cluster.
Interestingly, RS 6371 falls to the red of the RGB in the $V$ vs.\ $V-I$ CMD of
\citet{Stetson2003}, but appears to be a normal cluster giant in the $g'$ vs.\ $g'-r'$
CMD of \citet{Platais2011} and is thus not reported as an SSG in the Platais
et al. sample. 6371 is also a known photometric variable \citep[][V9]{deMarchi2007},
identified as an eclipsing binary within the \textit{Kepler} field, and also
an H$\alpha$ emitter \citep{vandenBerg2013}.

\paragraph{NGC 6819 } \citet{gosnell:12} identify 52004 with their X-ray source X9 in the intermediate-age ($\sim$2.4 Gyr) open cluster NGC 6819.  
The optical and IR photometry given in Table~\ref{sumtab} is from \citep{kalirai:01} and 2MASS, respectively.
This star has a proper-motion membership probability of 
99\% from \citet{platais:13}.  Though it has many spectroscopic observations from WIYN/Hydra, the source is a rapid rotator, and therefore reliable
radial velocities are difficult to obtain \citep{hole:09}. 
\citet{gosnell:12} find the X-ray and optical properties of this source to be consistent 
with an active binary, and note that it is similar to an RS CVn. 
The source is clearly in the SSG region in 
a $V$ vs.\ $B-V$ CMD (see Figure~\ref{f:CMDobs}), though it is not an obvious outlier in the ultra-violet
CMD presented in 
\citet[][their Figure 5b]{gosnell:12}.

\paragraph{NGC 7142 } \citet{sandquist:11} performed a photometric variability study of this intermediate age ($\sim$3.6 Gyr) open cluster, and 
discovered the source V4 to have variability on multiple timescales and amplitudes, including trends of $\sim$0.01 mag (particularly in $B$ and $V$)
in a given night, plus longer timescale variations over tens of days at about 0.5 mag (in $B$, $V$ and $R$).  
(We take the ID and optical photometry for this star from \citet{sandquist:11} and IR photometry from 2MASS.)
Investigation of their best-seeing images
does not indicate any binary companion.  They note that the location of V4 in the CMD is reminiscent of the SSGs in M67, though V4 has higher-amplitude 
photometric oscillations.  We place V4 in the RS category (see Figure~\ref{f:CMDobs}).  \citet{dias:14} find V4 to have a 91\% proper-motion membership probability.  
As also noted by \citet{sandquist:11}, radial-velocity observations for V4 would be very important to confirm cluster membership and 
investigate for a binary companion.  To our knowledge, there is no published X-ray survey of NGC 7142.

\subsection{Globular Cluster Observations}

\begin{figure}[!t]
\plotone{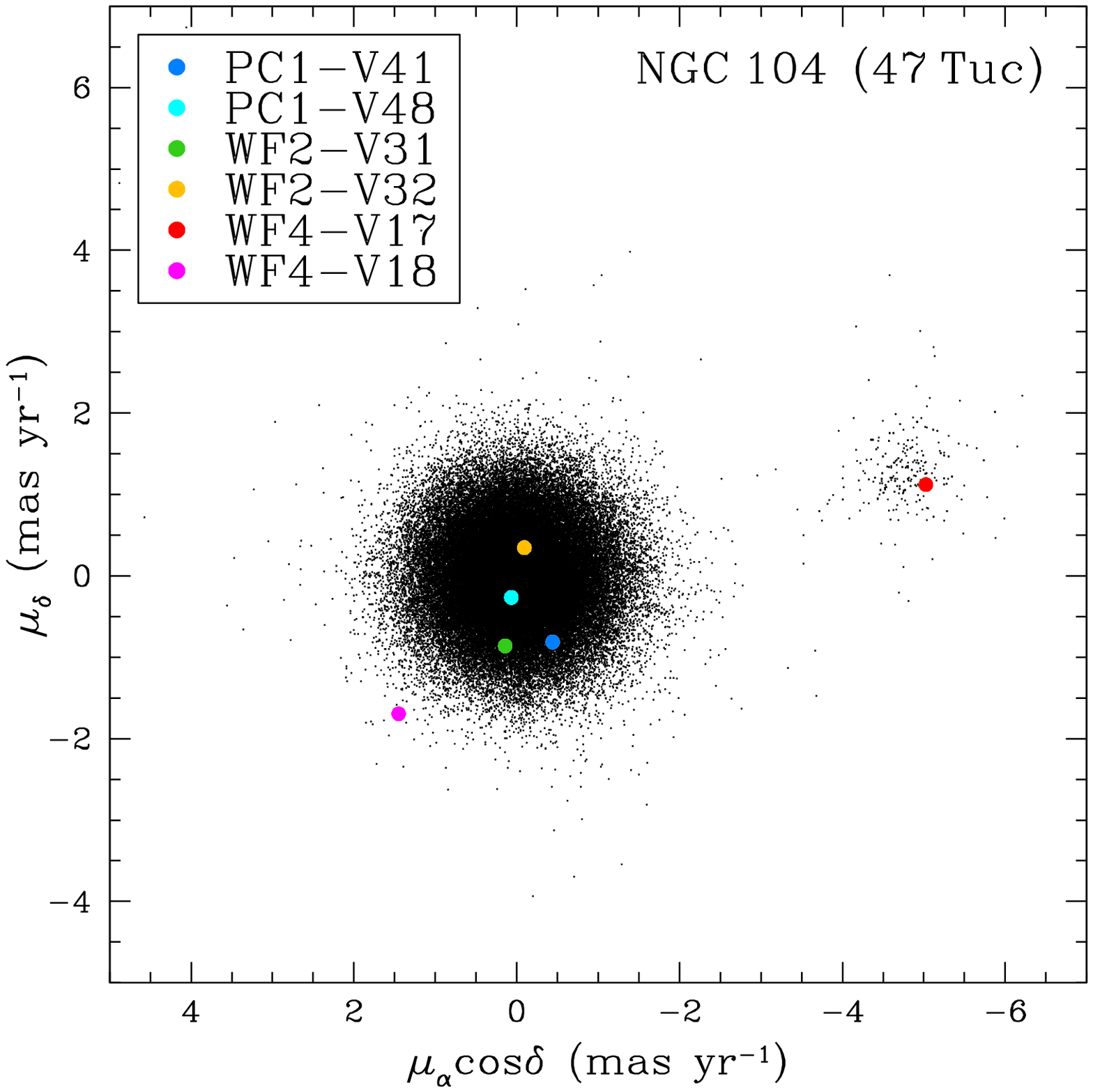}
\plotone{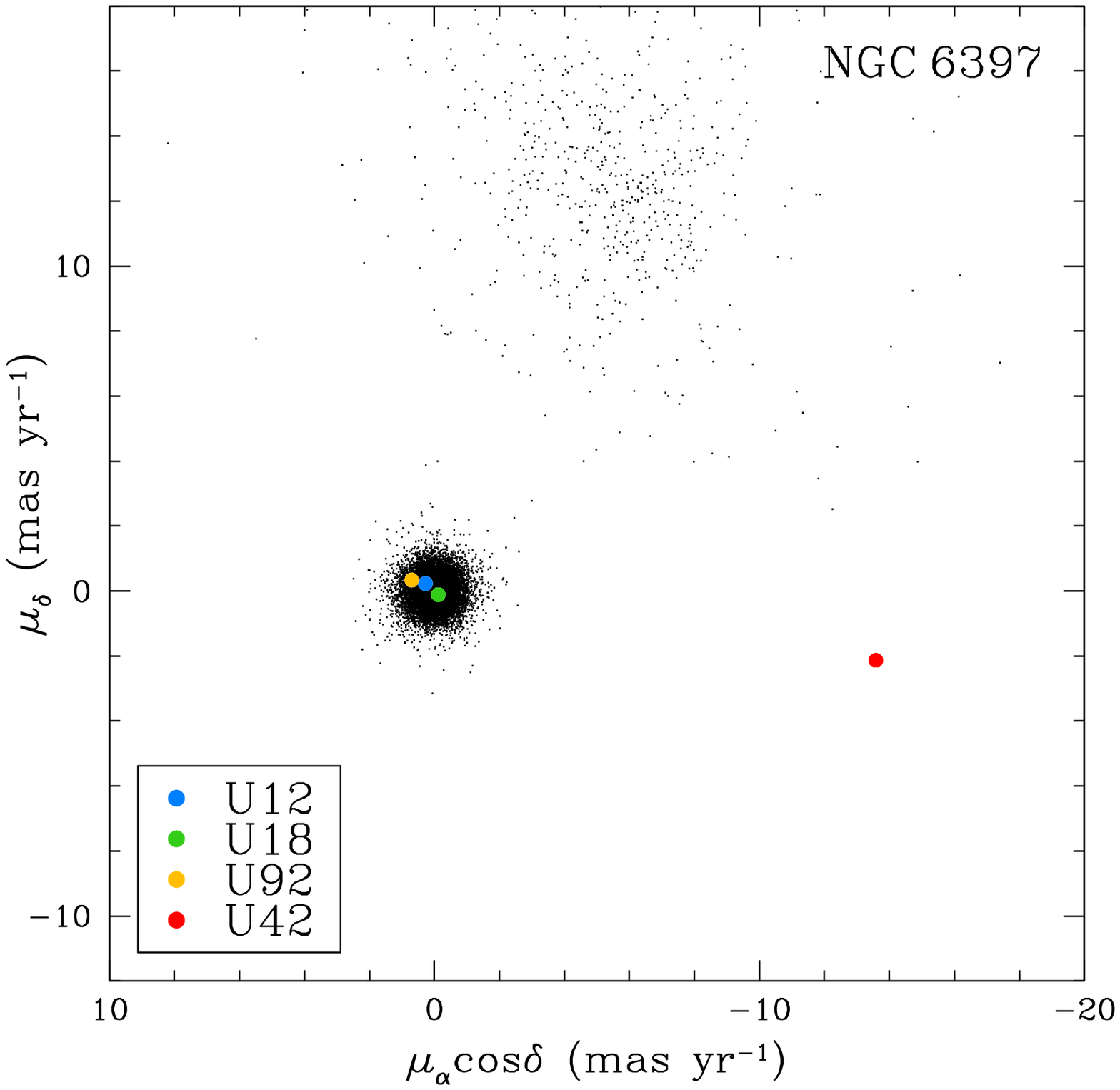}
\caption{
Proper-motion diagrams for NGC 104 (47 Tuc, top) and NGC 6397 (bottom). Stars from our sample are plotted in colored symbols, while the 
rest of the stars in the direction of each cluster, respectively, are plotted in black points.
Uncertainties on the proper-motion measurements for the stars in our SSG sample are smaller in size than the colored dots.
In both clusters, the members are easily distinguished visually, and confirmed through our more detailed analyses, 
as those having velocities consistent with the bulk motion of the cluster stars (where here the mean cluster motion is shifted to the origin).
Further details for these stars in both clusters are provided in the text of Section~\ref{s:obs}.
\label{f:HSTPROMO}
}
\end{figure}

\begin{figure*}[!t]
\begin{tabular}{cc}
\includegraphics[width=0.45\linewidth]{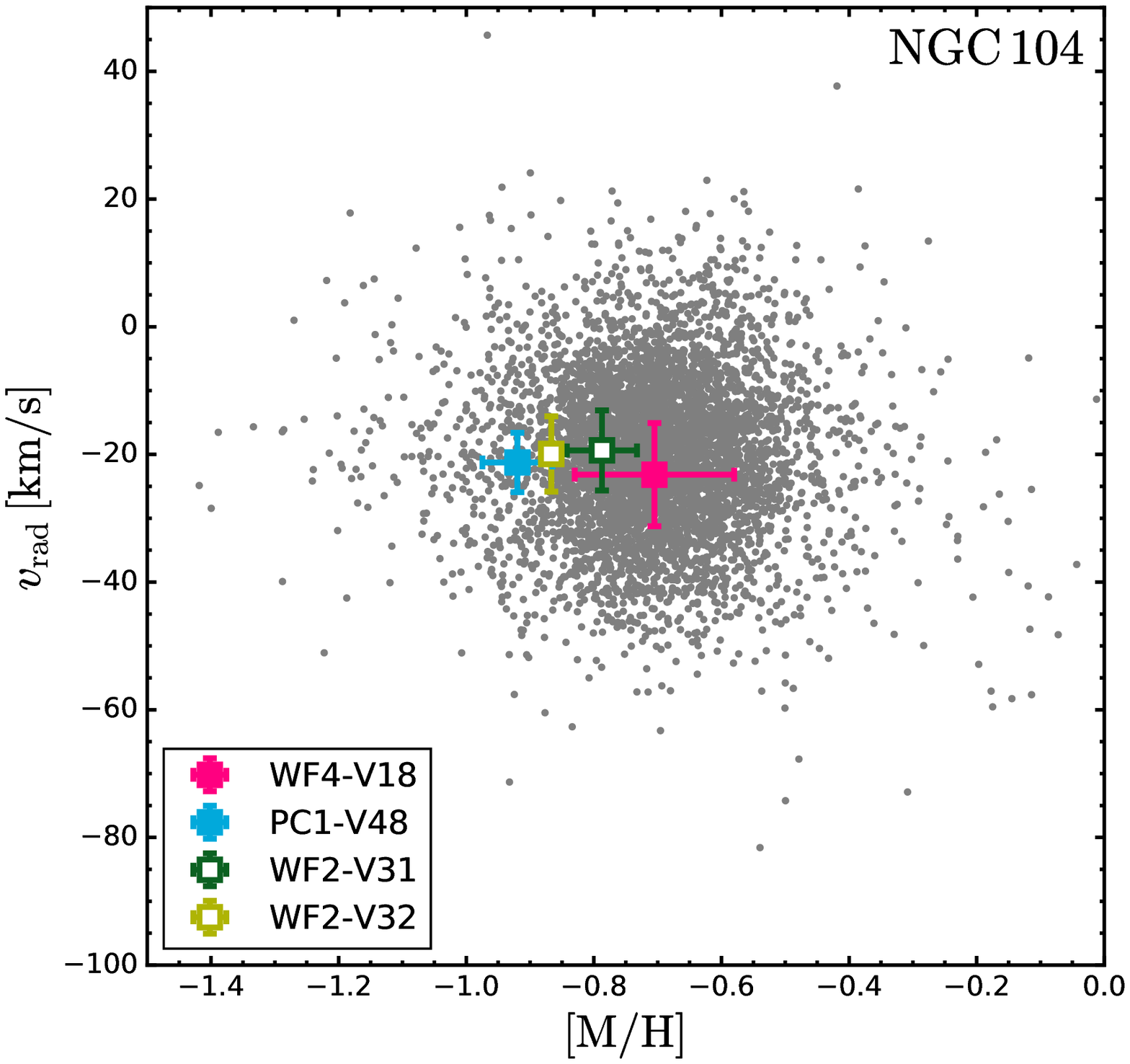} & \includegraphics[width=0.45\linewidth]{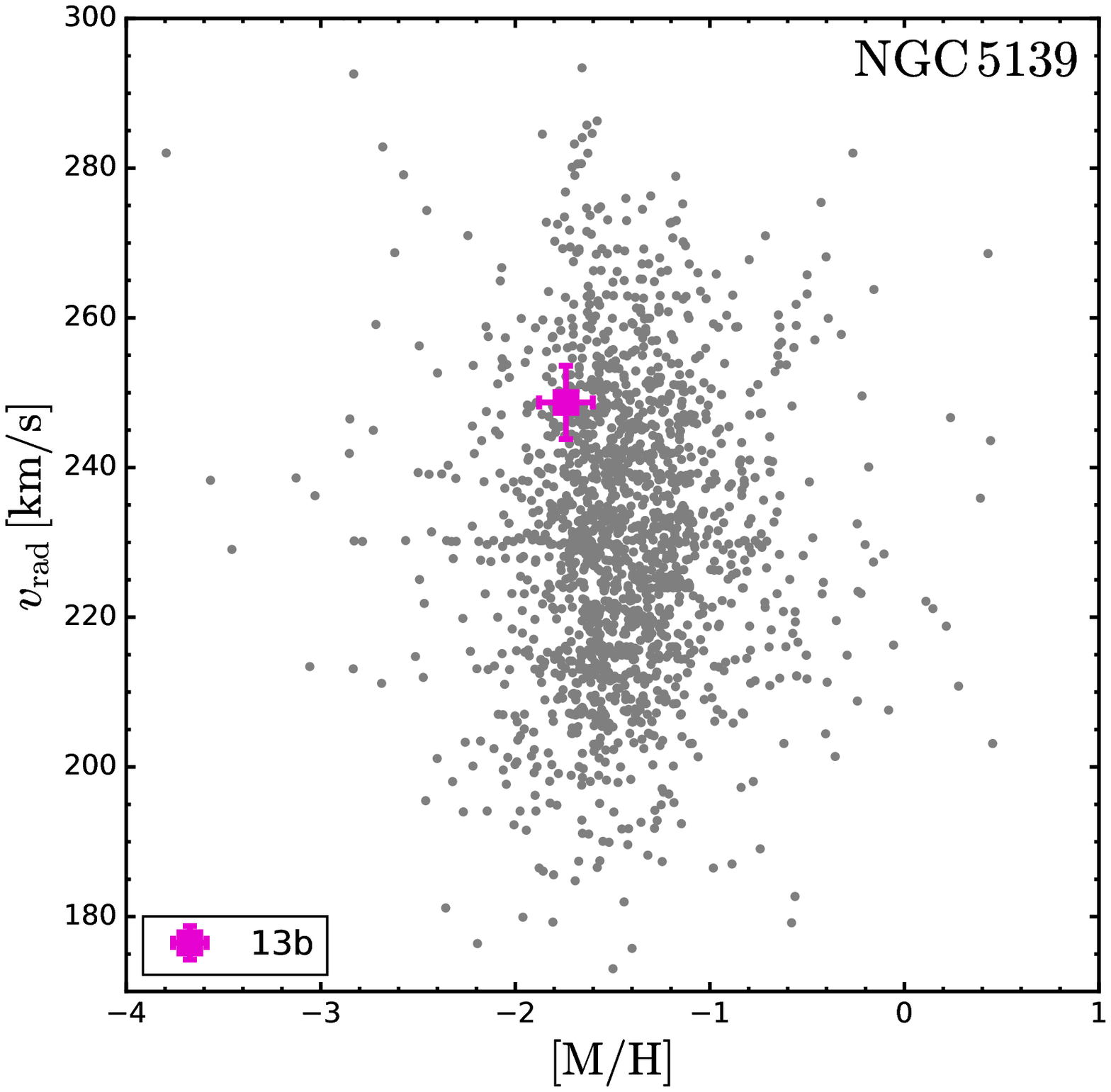} \\
\includegraphics[width=0.45\linewidth]{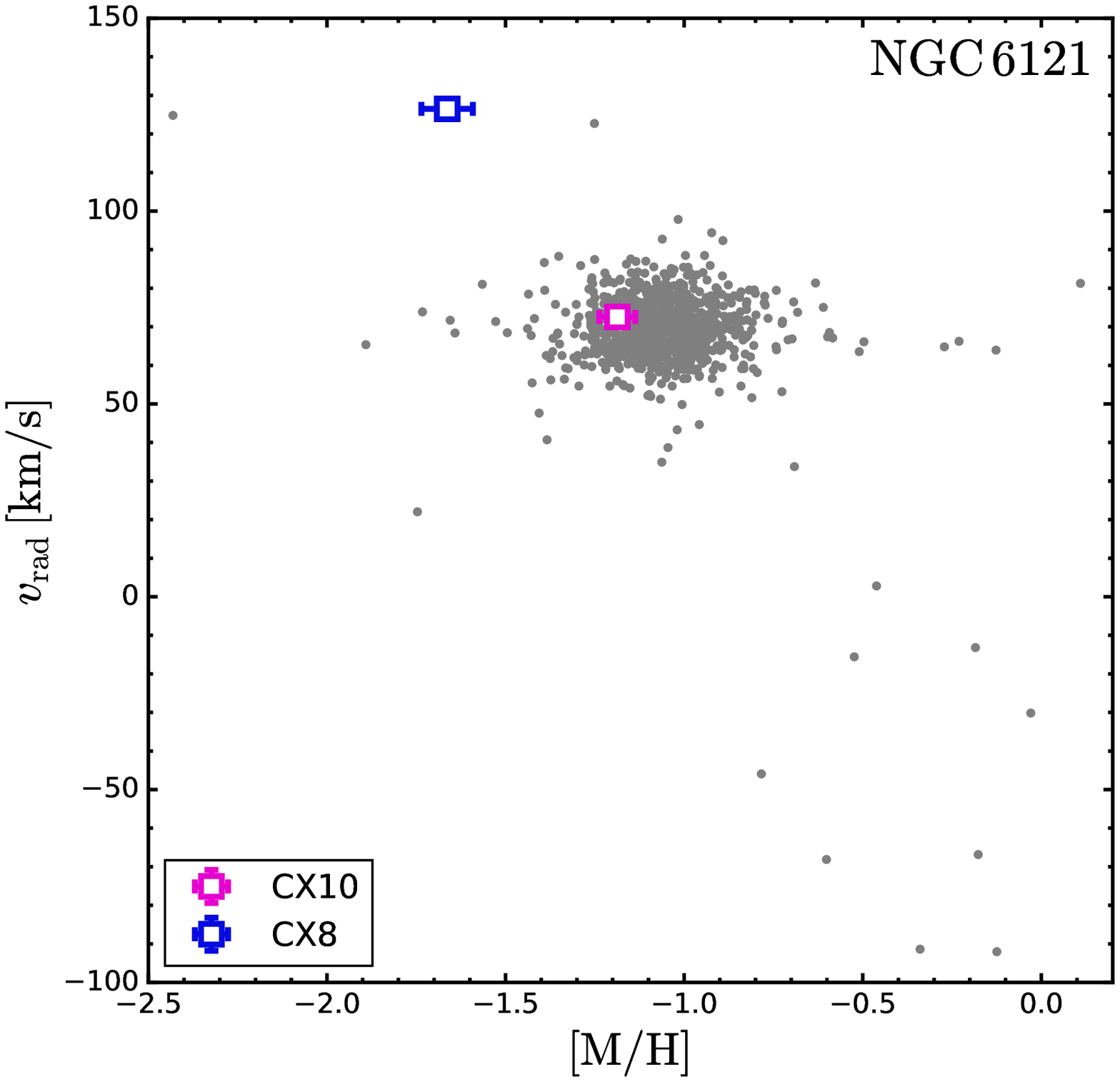} & \includegraphics[width=0.45\linewidth]{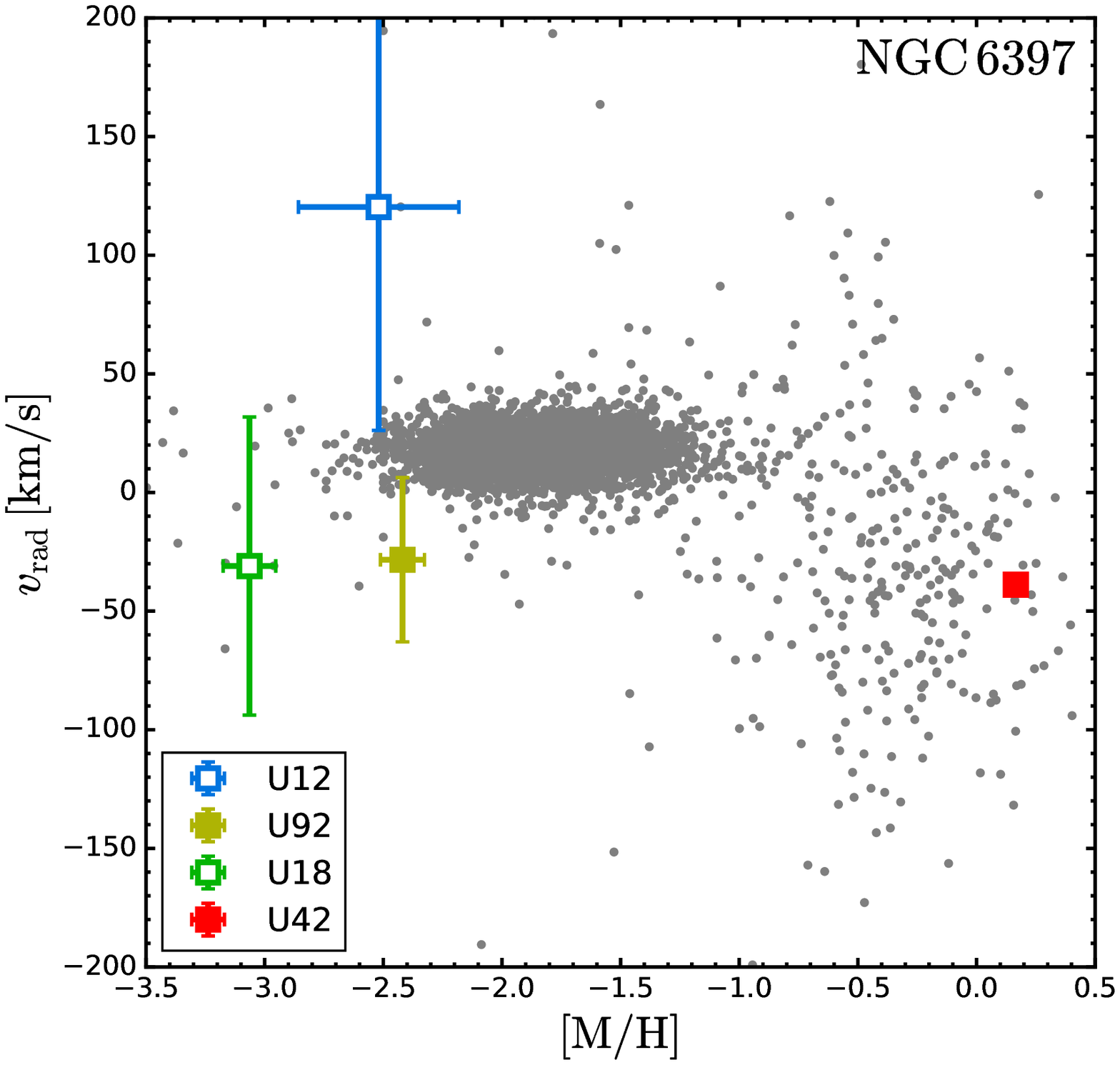} \\
\end{tabular}
\caption{
Radial-velocity ($v_{\rm rad}$) plotted against metallicity ([M/H]) for NGC 104 (47 Tuc, top left), NGC 5139 ($\omega$~Cen, top right), 
NGC 6121 (bottom left) and NGC 6397 (bottom right). 
Stars from our sample are plotted in colored symbols with error bars.  The rest of the stars surveyed for each cluster are shown in 
gray points.  Only stars with robust $v_{\rm rad}$ and [M/H] measurements are shown.  Vertical error bars in the plots 
for 47 Tuc, $\omega$~Cen and NGC 6397 account for the radial-velocity variability.  NGC 6121 has only one epoch of 
observations, and the vertical error bars show the (much smaller) uncertainties on individual measurements. Again, we caution that 
the single radial-velocity epoch for NGC 6121 may not show the true center-of-mass motion of binary stars
(as may be the case for CX8).
We use open symbols for stars that show significant H$\alpha$ emission in the MUSE spectra that likely affects our metallicity measurements.
Further details for all of these stars are provided in the text of Section~\ref{s:obs}.\\
\label{f:MUSE}
}
\end{figure*}

\paragraph{NGC 104 (47 Tuc) } Shortly after the discovery of the SSGs in M67, \citet{albrow:01} noted a population of six photometric variable stars that reside 
in the SSG region in a CMD of the globular cluster 47 Tuc (and called them ``red stragglers''). 
\citet{edmonds:03} later added four additional sources to this list in their analysis of \textit{Chandra} X-ray observations. 
Optical photometry for these stars in Table~\ref{sumtab} are converted to the ground-based system from the \textit{HST} magnitudes from \citet{albrow:01} and \citet{edmonds:03}.
We are able to match six of these ten sources to the HSTPROMO catalog\footnote{HSTPROMO draws from the ACS Survey for Globular Clusters, \url{http://www.astro.ufl.edu/~ata/public_hstgc/}} 
\citep[][and include their positions in Table~\ref{sumtab}, which have an average epoch of observations of 2006.2]{bellini:14}, 
and can therefore evaluate their proper-motion memberships. 
(The other sources were rejected from the proper-motion pipeline due to contamination from nearby stars or poor PSF fitting in one or more epochs, 
or they were simply outside of the field of view of the proper-motion catalog.)
After careful cleaning and analysis of the full cluster data set, we find that all but two of these six sources have relative velocities within 3$\sigma$ of the 
cluster's mean motion, and we therefore identify these four stars as proper-motion members.
WF4-V17 is clearly a proper-motion non-member of 47 Tuc, and is not included in our table (see Figure~\ref{f:HSTPROMO}, where the 
proper-motion errors for the six target stars are smaller than their colored symbols). Instead WF4-V17 is a red giant in the Small Magellanic Cloud. 
WF4-V18 has a 2D velocity 3.75$\sigma$ from the cluster mean, but also has a rather poor $\chi^2$ value for the linear fit defining its proper-motion 
\citep[see][]{bellini:14,watkins:15}, and also large uncertainties on the proper-motion.  
This source appears to be a MUSE radial-velocity member (see Figure~\ref{f:MUSE}), so we keep it in our table.  However, as with other sources with uncertain 
membership, we will not include WF4-V18 in further analyses.
Three more of these sources were also observed by the MUSE multi-epoch radial-velocity survey \citep{kamann:13,kamann:16}, who confirm their cluster membership 
based on both velocity and metallicity. Two of these sources show H$\alpha$ emission. 
The MUSE radial velocities indicate that all four of these sources show strong radial-velocity 
variability indicative of binary companions, with the strongest radial-velocity variable (WF2-V32) reaching an amplitude of $>$30 \kms.  
We currently do not have sufficient epochs of radial velocities to derive orbital solutions, and therefore center-of-mass radial velocities.  
Thus we do not quote radial-velocity membership probabilities for these stars in Table~\ref{sumtab}; we show their mean velocities, relative to 
the rest of the 47 Tuc MUSE sample in Figure~\ref{f:MUSE}.
Seven of these nine candidate cluster members are detected in X-rays \citep{grindlay:01,edmonds:03}, and
\citet{albrow:01} note that their X-ray luminosities are consistent with that expected for a chromospherically active subgiant star in an RS CVn type system.
Of additional interest, PC-V11 (also known as W36 in \citealt{edmonds:03} and AKO 9 \citealt{auriere:89}) 
is a known CV in the SSG region \citep[e.g.][]{grindlay:01,knigge:02}.

\paragraph{NGC 5139 ($\omega$~Centauri) }  $\omega$~Cen  has (at least) seven distinct sequences apparent in 
the optical/IR CMD \citep{villanova:07, bellini:10}, and also has very sensitive \textit{Chandra} imaging \citep{haggard:09,haggard:13}.
Of particular interest here, the ``anomalous RGB/SGB'' (sequence D from \citealt{villanova:07}, and also known as RGB/SGB-a, \citealt{lee:99,pancino:00,ferraro:04})
contains $\sim$10\% of the subgiant stars, 
has a subgiant branch that is significantly fainter than the other subgiant branches, and a red-giant branch that is significantly redder than the other red-giant branches.  
In other words, the anomalous RGB/SGB runs through the SSG and RS regions of the $\omega$~Cen CMD, relative to the bluer and brighter sequences.
\citet{villanova:07} find the anomalous RGB/SGB  to be old ($\sim$ 13 Gyr) and metal rich ([Fe/H] $\sim$ -1.1) in comparison to the other branches.
Interestingly, \citet{cool:13} discovered eight X-ray sources lying within the anomalous RGB/SGB on the CMD (with IDs containing numbers and a letter in Table~\ref{sumtab}). 
Three of these sources, all candidate SSGs (22e, 32f and 43c), are identified as H$\alpha$ ``Bright?'' by \citet{cool:13}.
We are able to match six of these eight sources to either the \citet{bellini:09} proper-motion study or the MUSE radial-velocity survey, all of which 
are consistent with cluster membership (though 32f has only a 43\% proper-motion membership probability). 
These \citet{cool:13} stars form a relatively tight sequence on the CMD.
($UBVRI$ magnitudes for the $\omega$~Cen stars in Table~\ref{sumtab} are from \citet{bellini:10}, where available, except, for the \citet{cool:13} sources,
where we convert their \textit{HST} magnitudes to ground-based $B$ and $R$ and use these instead.)
Similarly tight sequences of SSG and/or RS stars are not immediately apparent in other clusters 
(see Figure~\ref{f:CMDobs}; though the number of SSG and RS stars in most clusters is perhaps too small to discern a sequence in a CMD).
Without metallicity measurements, it is unclear whether these \citet{cool:13} X-ray sources are associated with the anomalous RGB/SGB or with some different branch.  
If they are associated with the anomalous RGB/SGB, 
then they may not be SSG (or RS) stars 
according to the CMD definition from Section~\ref{s:intro}
(unless all of the anomalous RGB/SGB are SSG and RS stars in relation to some different branch).  
However, in this scenario the anomalous RGB/SGB would have a factor of 5-15 times more X-ray sources (above the detection limits)
than the other branches that dominate the cluster mass, and would indicate a strong preference for X-ray sources at higher metallicities in $\omega$~Cen, 
a trend not observed elsewhere \citep{cool:13}.  
On the other hand, if these X-ray sources are instead associated with a different branch (by metallicity), then they would fit our definition 
of SSG and RS stars.
We choose to include them here, and future metallicity measurements for these stars will be very important to confirm their identity as SSG/RS stars.
Regardless of the nature of the \citet{cool:13} sources, \citet{rozyczka:12} identify an additional 13 3D kinematic cluster members in the SSG and RS regions 
in the cluster (not confined to the anomalous RGB/SGB).
(Note, SSG 
candidate 23 lies blueward of the $R$ vs.\ $B-R$ isochrone
in Figure~\ref{f:CMDobs}, but inside the SSG region in other filter combinations.)
This sample includes three radial-velocity variables plus four 
additional ``suspected radial-velocity variables'' (all of which we identify as ``var'' in Table~\ref{sumtab}), and one W UMa photometric variable.  
Furthermore, \citet{rozyczka:12} state that it is conceivable that the remaining 5 objects may also be thus far undetected binaries.

\paragraph{NGC 6121 (M4) } \citet{bassa:04} identify two candidate SSGs in their \textit{Chandra} X-ray survey of this globular cluster, CX8 and CX10.
For source CX8 we take the optical and IR photometry from \citet{stetson:14} and 2MASS, respectively, and for CX10 we convert the 
\textit{HST} photometry from \citet{bassa:04} to ground-based $V$ and $I$.
\citet{zloczewski:12} find both to be likely proper-motion members (each with a proper-motion $<2.5\sigma$ from the cluster mean).
Both CX8 and CX10 fall within the MUSE sample, but there is only one epoch of observations for this cluster (and therefore radial-velocity variability 
is unknown). 
CX10 appears to be a member by both radial velocity and metallicity (Figure~\ref{f:MUSE}).
CX8 appears separated from the cluster distribution.
Both CX8 and CX10 show H$\alpha$ emission in their spectra, which may bias the metallicity measurements.
Also, if CX8 is a binary, the one radial-velocity epoch may not reflect the center-of-mass motion.
\citet{bassa:04} note that CX8 coincides with the photometric variable V52 from \citet{kaluzny:97}, which they classify as a BY Dra system 
(generally thought to contain MS stars with variability arising from spots and chromospheric activity) with a period of $\sim$0.78 days.
\citet{kaluzny:13} continue to monitor V52, and note that the periodicity remains coherent over 14 years, which they take as indication that 
V52 is a binary star.  However, contrary to \citet{bassa:04}, \citet{kaluzny:13} specifically state that none of the X-ray sources from \citet{bassa:04} coincide with V52
(which indeed they find puzzling, given the expected chromospheric activity of such a star).  
\citet{nascimbeni:14} confirm the $\sim$0.78 day photometric period, and also associate this star with CX8 (and with their 7864).  They categorize this 
source as ``unclassified/uncertain''.
We choose to provide the photometric period for V52 in Table~\ref{sumtab} as related to CX8.
Importantly, based on radio observations, Strader et al.\ (in prep) suggest that CX8 has a compact object companion, with a high likelihood that the companion is a black hole.
Due to the uncertainty in binarity (and hence the center-of-mass radial velocity) and the 
probable bias in the metallicity measurement, we do not suggest a radial-velocity membership for CX8.
Finally, for completeness, we note that \citet{bassa:04} identify another source, CX24, that also falls to the red of the standard MS but is fainter than the typical 
SSG region as defined here. The optical counterpart to CX24 varies by more than 1 mag in brightness, between their \textit{HST} observation dates.
\citet{bassa:04} suggest that CX24 is a foreground object, and we therefore do not include this source in our table.

\paragraph{NGC 6218 (M12) } \citet{lu:09} identify an X-ray source (their CX2) in NGC 6218 with a ``relatively hard X-ray color'', for which they find three potential 
optical counterparts.  One of these potential optical counterparts, CX2b, falls in the SSG region on an optical CMD, 
and we convert their \textit{HST} photometry for this source to ground-based $B$, $V$ and $R$ for Table~\ref{sumtab}.
(The other two fall on, or possibly to the red depending 
on the color choice, of the main-sequence, and to the blue of the main sequence, respectively.)  \citet{zloczewski:12} find this source to be $>2\sigma$ from the mean proper motion
of the cluster, and therefore categorize this star as a non-member.  We include this star in our table, though as with other similar sources, we 
will not include this star in our subsequent analysis.

\paragraph{NGC 6366 } \citet{bassa:08} identify one candidate SSG, CX5, in their \textit{Chandra} X-ray survey of this globular cluster.
To our knowledge there is no proper-motion membership probability available for this source in the literature, but \citet{bassa:08} conclude that this 
is a probable cluster member, based on the observed X-ray luminosity and their optical photometry 
(which we provide in Table~\ref{sumtab}).

\paragraph{NGC 6397 } Four \textit{Chandra} X-ray sources from \citet[][U12, U18, U42 and U92]{cohn:10}, 
reside in the cluster SSG region.
(We convert the \citet{cohn:10} \textit{HST} magnitudes to ground-based $B$ and $R$ for Table~\ref{sumtab}.)
All of these sources are found in the HSTPROMO catalog, and we use the HSTPROMO positions in Table~\ref{sumtab}, which have
an average epoch of observations of 2006.4.
All but one of these sources are probable members from both HSTPROMO proper motions and MUSE radial velocities and metallicites.
The remaining source, U42, falls well outside of the cluster distribution (see Figures~\ref{f:HSTPROMO}~and~\ref{f:MUSE}). 
We therefore exclude U42 from our sample (and also note that U42 is somewhat redder than most of the SSGs in our sample).
All three members are photometric variables \citep{kaluzny:06}, 
and all show remarkably high amplitude radial-velocity variability. 
Each were observed twice within about 24 hours by the MUSE team and have radial velocities that differ by 70 to 200 \kms.  
Importantly, U12, whose radial velocity was observed to vary by 200 \kms\ in a day, is a known millisecond pulsar 
\citep[MSP;][]{damico:01,ferraro:03}, and \citet{bogdanov:10} suggest that U18 is also a MSP.  
\citet{kaluzny:06} attribute the short-period photometric variability for both of these sources (U12 = V16; U18 = V31) to ellipsoidal variations 
(though they are somewhat uncertain about that characterization for U18).
MUSE spectra show that both U12 and U18 have H$\alpha$ in emission.
Finally, source U92 (V7) is an eclipsing W UMa binary, and shows H$\alpha$ in absorption in the MUSE spectra.
For completeness, we also note that U63, U65, U86 are located redward of the MS but fainter than the SSG region discussed here.
All are X-ray sources, and U65 and U86 each have an H$\alpha$ excess.
However, \citet{cohn:10} show that these three stars have proper motions that are inconsistent with both the cluster and field distributions, and therefore 
their membership status is unknown.  They suggest these three stars may be foreground active binaries, and we do not include them in our sample (or Table~\ref{sumtab}).

\paragraph{NGC 6652 } Source B is one of two known LMXBs in the globular cluster NGC 6652 \citep{heinke:01}, and was studied in detail by \citet{coomber:11} 
and \citet{engel:12}.  
This source ``flickers'' on timescales less than 75 seconds (the exposure time for the \citealt{engel:12} observations), with amplitudes of  $\sim1$ mag in g' and $\sim$0.5 mag in r',
and therefore the optical photometry, and particularly an optical color, is highly uncertain.
In Table~\ref{sumtab}, we provide an estimate of the $V$ magnitude from \citet{heinke:01} for reference.
\citet{engel:12} suggest a color potentially redder than the MS, akin to the SSGs in other clusters.
\citet{deutsch:00} find a possible photometric period of 43.6 minutes, though \citet{heinke:01}
and \citet{engel:12} suggest that this period is spurious.
The source also flares in X-rays on timescales down to 100s, and can range from $L_X$(0.5-10.0keV)$ < 2 \times 10^{33}$ erg s$^{-1}$ up to 
$L_X$(0.5-10.0keV)$ \sim 10^{35}$ erg s$^{-1}$ with no detectable periodicity, though its long-term $L_X$ is observed to be relatively constant 
since 1994 \textit{ROSAT} observations \citep{coomber:11}.
The high peak in $L_X$ is strong evidence for a neutron star or black hole companion. However, the variability and somewhat low X-ray luminosity 
is unusual for typical LMXBs. Because the optical photometry is so uncertain, we cannot reliably classify this source as an SSG or RS; we include this 
source in Table~\ref{sumtab} for reference, but do not include it in our subsequent analysis. 

\paragraph{NGC 6752 } \citet{kaluzny:09} find three photometric variables in the SSG region in this globular cluster.  
(IDs and photometry for these three stars in Table~\ref{sumtab} are from \citealt{kaluzny:09}).
Two are roughly the same $V$ magnitude as the base of the giant branch (V19 and V20), while the other is fainter.  
V19 is quite red, somewhat similar in this regard to U42 in 
NGC 6397 (which appears to be a non-member). 
\citet{kaluzny:09} propose that the photometric variability for these sources is due to binarity and possibly ellipsoidal variations, with 
a degenerate companion.  To our knowledge these sources are not detected in X-rays.  \citet{zloczewski:12} find that V19 and V20 are $>2\sigma$ from the
mean of the cluster proper-motion distribution, and categorize them as non-members.  Again, as this is below our $3\sigma$ limit, we choose to include these 
stars in our table as candidates, but we will not include them in our subsequent analysis.

\paragraph{NGC 6809 (M55) }  We find three candidate SSGs from the literature in this globular cluster, CX7 and CX29 from \citet{bassa:08} and V64 from \citet{kaluzny:10}.
Both CX7 and V64 are cluster members, while CX29 is likely a non-member (at $>3\sigma$), from proper-motion measurements \citep{zloczewski:11}, and 
is therefore not included in our sample.
(In Table~\ref{sumtab}, the IDs and photometry for CX7 are from \citealt{bassa:08}, while the ID and photometry for V64 are 
from \citealt{kaluzny:10}.)
CX7 is detected in X-rays by \citet{bassa:08},  who note that this source is likely a magnetically active binary that has no significant H$\alpha$ emission.
CX7 is also a photometric variable from \citet[][V65]{kaluzny:10}, who posit that the photometric variability is either due to a contact binary observed 
at low inclination, or ellipsoidal variations suggesting a degenerate companion.  Moreover, \citet{kaluzny:10} suggest that the coherence of the photometric 
variations for both V64 and V65 (CX7) are indicative of binary companions.  
\citet{lane:11} find that CX7 is a cluster non-member from radial-velocity observations, but given the evidence for binarity and therefore the unknown 
center-of-mass velocity (without an orbital solution), we suggest that the radial-velocity membership status is uncertain.

\paragraph{NGC 6838 (M71) } \citet{huang:10} identify two SSG candidates in this globular cluster.
Their source s02 has an X-ray to optical flux ratio indicative of an active binary, and X-ray variability 
that likely indicates flaring from a chromospherically active star \citep{elsner:08}. The X-ray source s19 has three potential 
\textit{HST} optical counterparts within the \textit{Chandra} error circle.  All three of these potential counterparts would have 
X-ray to optical flux ratios consistent with active binaries (though s19c could also be interpreted as a CV).  
\citet{huang:10} suggest that s19a is the most likely counterpart to the source, evidently because this would place the star in 
the SSG region.  
Photometry for both NGC 6838 sources are converted from the \citet{huang:10} \textit{HST} filters to ground-based magnitudes for Table~\ref{sumtab}.

\subsection{\textit{Kepler} ``No-Man's-Land'' Stars}
\label{ss:kepler}

\citet{batalha:13} and then \citet{huber:14} identify a subset of 
roughly 5\% of the \textit{Kepler} targets (nearly 10,000 stars) with (photometric) surface gravities and temperatures that are 
inconsistent with the expectations for normal stars from standard isochrones.  More specifically, these are generally stars of G or K spectral type that have surfaces 
gravities that are too high and temperatures that are too cool to be consistent with any isochrone less than 14 Gyr 
(even at extremely high metallicities, e.g., see Figure 14 in \citealt{huber:14}).  Moreover, these stars fall in the 
SSG and RS regions.  Almost definitely some of these stars simply have incorrect surface gravities and/or temperatures. 
However, \citet{huber:14} follow up a subset of these stars with spectroscopic classifications 
from the SEGUE catalog \citep{yanny:09}, and find that even these more accurate surface gravities and temperatures do not move 
all stars out of the ``No-Man's-Land'' regime.  Indeed they state that ``a considerable number of SEGUE classifications remain in the 'No-Man’s-Land' zone''.
We suggest that there may be a substantial population of \textit{field} SSG stars within the \textit{Kepler} ``No-Man's-Land'' sample.  

If even a subset of the ``No-Man's Land'' stars are indeed SSGs, then their discovery in the field suggests that 
SSGs can form through channels that do not require dynamical encounters within star clusters. 
(This may not be surprising, since it is also well known that blue stragglers exist in clusters and the field, 
and can form through mechanisms mediated by dynamics as well as through isolated binary evolution.)
We point out these stars here to motivate further observations and analyses that might help confirm whether or not these stars are indeed SSGs.

\subsection{Summary}

In summary, we compile a sample of 65 stars in 16 star clusters identified in the literature as either SSG or RS stars.  We classify 56 of these stars as SSGs, based on 
our CMD definition described above (and shown in Figure~\ref{f:CMDobs}).  In the following sections, we describe and attempt to characterize the biases and incompleteness in 
this sample, and we discuss the cluster membership status of these stars.  From our analysis presented in the following sections, considering the proper motions, radial velocities, 
photometric variability and X-ray luminosities, we conclude that the vast majority of these stars are indeed cluster members.  We select these highly likely 
cluster members when investigating the SSG demographics in Section~\ref{s:agg}.

\section{Observational Biases in the Sample}
\label{s:bias}

We do not attempt to formally correct for the selection effects or incompleteness that is likely present in the sample of SSG/RS stars in Table~\ref{sumtab}.
We will, however, limit their impact in the analysis of these data by including only the most likely cluster members.

Most of the sources from the open clusters listed in Table~\ref{sumtab} were identified from 
comprehensive radial-velocity and proper-motion membership surveys.  Most of the globular cluster sources, on the other 
hand were, initially identified in X-ray surveys (without comprehensive membership surveys).

The detection limit of most of these X-ray surveys is of the order of 10$^{30-31}$ erg s$^{-1}$, which 
appears to be the characteristic X-ray luminosity of these SSG stars.
Therefore, (a) there may be more SSG/RS stars in clusters with X-ray luminosities below $\sim$10$^{30-31}$ erg s$^{-1}$ that 
have not been identified in the literature, and (b) there may be unidentified SSG/RS stars in  
clusters that currently have less sensitive X-ray observations, not reaching 10$^{30-31}$ erg s$^{-1}$.
The large frequency of X-ray emitting SSG and RS stars may be simply due to the discovery method, although 
the open cluster sample suggests otherwise.  
Proper-motion surveys of a large sample of globular clusters are nearing completion, which will help 
to identify non-X-ray-detected SSG/RS stars (and help to further eliminate non-members from our SSG/RS sample).

X-ray surveys of globular clusters have often targeted the most dynamically active clusters, since observations indicate a trend of 
increasing frequency of X-ray sources with increasing collision rate \citep{pooley:03,bahramian:13}.  Therefore the globular cluster sample here is 
likely biased toward the more massive and dense clusters.  
Indeed, NGC 6397 and NGC 6752 are core-collapsed clusters in the \citet{harris:10} catalog.
Open clusters that have a particularly large number of stars are also often selected for radial-velocity, photometric, and X-ray surveys, 
and therefore again, our sample of open clusters is likely biased toward the most massive clusters at a given age.  

Also, not all sources listed in Table~\ref{sumtab} were monitored for photometric variability, and not all sources (particularly those in the globular clusters) were 
observed for radial-velocity variability.  Therefore the frequency of variables in our sample is a lower limit. 

Furthermore, as with most studies of periodic data, we expect that the ability to detect periodicity in this sample decreases
with increasing period.  Therefore the period distribution of the known variable SSG/RS stars (e.g., Figure~\ref{f:per}) may be biased toward shorter periods.  

Lastly, as these clusters are all at different distances and have been observed using different telescopes and instruments, the radial coverage of the 
clusters varies across our sample.  We show the maximum radial extent ($r/r_{\rm c}$) of the relevant SSG discovery survey for each cluster in Table~\ref{clustab}.
The cluster with the smallest radial coverage that contains SSG stars is NGC 6121 (one of the closest globular clusters 
in this group), with a maximal radial coverage of 3.3 core radii;
note that NGC 7142, with a smaller coverage, does not have any SSG stars.
 When necessary, to reduce the effect of radial incompleteness amongst our surveys, 
we will limit our sample to only include those SSGs within 3.3 core radii from their respective clusters for the analysis in Section~\ref{s:agg}. 

\section{Evidence for Cluster Membership}
\label{s:contam}

\begin{deluxetable*}{lcccccccccc}
\tabletypesize{\tiny}
\tablecaption{Photometric Field Contaminant Probabilities \label{obstab}}
\tablehead{\colhead{Cluster} & \colhead{$N_\text{o}$} &  \colhead{$N_\text{oV}$} & \colhead{$N_\text{oX}$} &\colhead{$N_\text{e}$} & \colhead{$N_\text{eV}$} & \colhead{$N_\text{eX}$} & \colhead{$P$} & \colhead{$P_\text{V}$} & \colhead{$P_\text{X}$} \\
\colhead{} & \colhead{} &  \colhead{} & \colhead{} &\colhead{} & \colhead{} & \colhead{} & \colhead{[\%]} & \colhead{[\%]} & \colhead{[\%]} }
\tablewidth{0pt}
\startdata
&&&&&&&&&\\
\multicolumn{9}{l}{\textbf{Open Clusters}} \\
&&&&&&&&&\\
NGC 188                   &    3 &    2 &    1 &   12 &  0.251 &  0.099 &  99.9 &     2.7 &     9.4 \\
NGC 2158                  &    5 &    4 &    0 &   90 &  1.879 &  0.743 & 100.0 &    12.2 & \nodata \\
NGC 2682                  &    2 &    1 &    2 &   50 &  1.044 &  0.396 & 100.0 &    64.8 &     6.1 \\
NGC 6791                  &    5 &    4 &    4 &  253 &  5.281 &  2.089 & 100.0 &    77.2 &    15.9 \\
NGC 6819                  &    1 &    0 &    1 &  331 &  6.910 &  2.419 & 100.0 & \nodata &    91.1 \\
Total(Open Clusters)      &   16 &   11 &    8 &  736 & 15.364 &  5.747 & 100.0 &    89.8 &    22.2 \\
&&&&&&&&&\\
\multicolumn{9}{l}{\textbf{Globular Clusters}} \\
&&&&&&&&&\\
NGC 104                   &    7 &    5 &    6 &    0 &  0.000 &  0.000 &   0.0 &     0.0 &     0.0 \\
NGC 5139                  &   19 &    1 &    7 &   28 &  0.585 &  0.231 &  97.0 &    44.3 &     0.0 \\
NGC 6121                  &    2 &    1 &    2 &    0 &  0.000 &  0.000 &   0.0 &     0.0 &     0.0 \\
NGC 6366                  &    1 &    0 &    1 &   20 &  0.342 &  0.017 & 100.0 & \nodata &     1.6 \\
NGC 6397                  &    3 &    3 &    3 &    1 &  0.021 &  0.008 &   8.0 &     0.0 &     0.0 \\
NGC 6752                  &    1 &    1 &    0 &   30 &  0.626 &  0.182 & 100.0 &    46.5 & \nodata \\
NGC 6809                  &    1 &    1 &    0 &    6 &  0.125 &  0.041 &  99.8 &    11.8 & \nodata \\
NGC 6838                  &    2 &    0 &    2 &    1 &  0.021 &  0.008 &  26.4 & \nodata &     0.0 \\
Total(Globular Clusters)  &   36 &   12 &   21 &   86 &  1.720 &  0.487 & 100.0 &     0.0 &     0.0 \\
\enddata
\tablenotetext{}{Note: The expected numbers of sources do not account for kinematic information.}
\end{deluxetable*}

\subsection{Probability of Field Star Contamination}

The majority of the SSGs in our sample have proper-motion measurements indicative of cluster membership.  
Many of the sources also have radial-velocity measurements indicative of cluster membership.
However, the quality of these kinematic data varies between clusters, and eleven of our sources do not have any kinematic membership data.  
Therefore in this section, we first consider other membership indicators by examining the number of expected field stars
within the SSG region in each cluster CMD (Table~\ref{obstab}). 
We will then include kinematic membership information for the individual SSG/RS stars to provide an estimate of the probability that 
each SSG or RS star is a field star (Equation~\ref{e:field} and $P_\text{field}$ in Table~\ref{sumtab}).

First, active galactic nuclei (AGN) are a well known contaminant in X-ray surveys of star clusters. 
However, to our knowledge none of these SSGs are noted as extended sources in the literature.  Also, AGN are known to show non-periodic, stochastic flux variations on 
timescales of months to years \citep[see e.g., ][and references therein]{simm:15}, which is inconsistent with the photometric variability seen for the SSGs in our sample.
Therefore here we investigate stellar contaminants.  Specifically we investigate the probability that we would observe any galactic field stars, field-star X-ray sources or 
short-period stellar photometric variables, respectively, in the SSG region of the CMD in each cluster. 

We utilize star fields from the Besan\c{c}on model of the Milky Way \citep{robin:03} within the maximum survey radius for each cluster (see Table~\ref{clustab}), 
that contains SSGs in Table~\ref{sumtab}.  
We use the same color and magnitude combination for each cluster as shown in Figure~\ref{f:CMDobs}, respectively, and identify an approximate region around the observed SSGs that 
extends from the bluest portion of the region shown in Figure~\ref{f:CMDobs} to the color of the reddest SSG in the cluster plus 0.05, and the magnitude of the faintest SSG 
in the cluster plus 0.5. 
(We also perform a similar analysis for the RS stars, for Equation~\ref{e:field} below, limiting the region to the color of the reddest RS in the cluster plus 0.05 and 
to the magnitude of the brightest RS in a cluster minus 0.5.)
These offsets in color and magnitude are somewhat arbitrary, but do not significantly affect the results presented here.
For each cluster region, we then count the number of field stars expected to fall in the SSG region of the appropriate CMD, and give this number as $N_e$ in Table~\ref{obstab}.
About 93\% of these expected field stars are dwarfs, and 94\% are spectral types G, K or M.  

We do not account for kinematic information in this analysis in Table~\ref{obstab} (which is partly responsible for the large numbers of stars predicted for the open clusters), 
whereas the SSG and RS stars are mostly drawn from samples of known kinematic cluster members.  The expected numbers of Galactic field stars given in Table~\ref{obstab} 
are upper limits.

We estimate the number of short-period binaries expected to be in the SSG region for each cluster, $N_{eV}$ in Table~\ref{obstab}, following the results from \citet{raghavan:10}.  
More specifically, we assume the field binary-star fraction 
is a function of spectral type from their Figure~12, and assume all binaries follow the same log-normal period distribution as the solar-type stars
(peaked at a mean value of $\log P = 5.03$ and with $\sigma_{\log P} = 2.28$, defining $P$ in days).  
This orbital period distribution should serve adequately for these foreground GKM dwarfs.
The log-normal period distribution predicts about 4\% of binaries should have 
orbital periods $<$15 days.  We then use this percentage and the appropriate binary fractions to estimate the numbers of expected binaries with periods $<$15 days in the 
SSG region of the CMD for each respective cluster, and give these numbers, $N_{eV}$, in Table~\ref{obstab}.
Note that this is an overestimate of the number of binaries that would be detected, as we have not accounted for the expected random inclinations, the 
potential for low-mass companions below a detection limit, observing cadence, etc.

Next we estimate the number of stellar X-ray sources expected for each cluster, $N_{eX}$ in Table~\ref{obstab}. 
Stellar X-ray sources can be either single or binary stars.  However, single X-ray emitters with luminosities of $10^{30} - 10^{31}$ erg s$^{-1}$ have ages $\lesssim$10 Myr
\citep[see ][and particularly their Figure~4]{preibisch:05}.  The Besan\c{c}on model predicts no stars of this young age in the fields of any of the clusters in this study.
RS CVn and BY Dra active binaries also have similar X-ray and optical properties as the SSGs (Figure~\ref{f:xray}).
\citet{eker:08} provide a catalog of known chromospherically active binaries.  From their catalog, we find that the BY Dra systems that have 
X-ray luminosities of order $10^{30} - 10^{31}$ erg s$^{-1}$ all contain GK dwarfs and have orbital periods between about 0.5 and 4 days.  RS CVn systems within this 
X-ray luminosity range contain primarily GK subgiant and giant stars and have orbital periods between about 0.5 and 100 days.  
If we consider only the portion of the \citet{raghavan:10} period distribution that would be occupied by BY Dra and RS CVn binaries (and assume that this same distribution holds for evolved stars), 
we find that about 1\% and 4\% of (field) stars should have appropriate orbital periods, respectively. Applying these percentages and the cuts in spectral types and luminosity classes
to the Besan\c{c}on model, we find a total of about six field X-ray active binaries expected in the directions of all open clusters studies here, and a total of less than one in 
the directions of all of the globular clusters in our sample.  Nearly all of the expected active binaries are BY Dra systems.

Also in Table~\ref{obstab}, we provide the observed numbers of SSGs ($N_o$, i.e., those with at least one color-magnitude combination that resides in the SSG region 
and do not have ``NM?'' in Table~\ref{sumtab}), 
the number of observed SSGs in binaries with orbital periods $<$15 days ($N_{oV}$, and here we assume that any short-period photometric variability 
is due to a binary companion of similar orbital period), and the number of observed SSG X-ray sources with luminosities of order $10^{30} - 10^{31}$ erg s$^{-1}$ ($N_{oX}$). 

We then calculate the cumulative Poisson probability, $P$, of observing at least $N_o$ SSGs (for those with at least 1 SSG) with an expected number of $N_e$ 
(found from the Besan\c{c}on model):
\begin{equation} \label{e:poisson}
P = 1 - e^{-\left(N_e\right)}\sum_{x=0}^{N_o-1} \frac{\left(N_e\right)^x}{x!} ,
\end{equation}
We also calculate the probability, $P_{V}$, that we would detect at least $N_{oV}$ short-period binaries in the SSG region when $N_{eV}$ are expected
(by replacing $N_o$ and $N_e$ in Equation~\ref{e:poisson} by $N_{oV}$ and $N_{eV}$, respectively).
Similarly we calculate the probability, $P_{X}$, that we would detect at least $N_{oX}$ X-ray sources in the SSG region when $N_{eX}$ are expected
(by replacing $N_o$ and $N_e$ in Equation~\ref{e:poisson} by $N_{oX}$ and $N_{eX}$, respectively).
We provide all of these probabilities in Table~\ref{obstab}, including total probabilities, considering the summed number of each type of star in the calculations, 
for the open and globular clusters respectively.  

Without any constraints on X-ray emission or binarity (or kinematic membership), the number of expected stars in the SSG region of the CMD, $N_e$, is similar to or greater than the number 
of true SSGs observed, $N_o$, in most clusters, and therefore the probability of observing 
at least $N_o$ stars in this region of the CMD in most clusters
($P$) is high. 
For some clusters, and especially for the globular clusters, taking also the photometric and radial-velocity variability (i.e., binarity) into account 
is enough to provide a high confidence level that the SSGs are not all field stars.
For the entire globular cluster sample we find a 0\% probability ($P_V$) that 
we would observe $N_{oV} = 12$ or more stars when $N_{eV}\sim 2$ stars are expected.
Therefore, even without accounting for kinematic membership information, it is exceedingly unlikely that all
variable SSGs in the globular clusters could be field stars.

The X-ray sources provide an even more stringent constraint.  
Each respective globular cluster is predicted to have $N_{eX}<1$ field X-ray source with an X-ray luminosity between $10^{30}$ and $10^{31}$ erg s$^{-1}$ in the SSG CMD region 
(considering both single and binary X-ray sources).  For every globular cluster, 
respectively,  we find that it is extremely unlikely that we would observe $N_{oX}$ X-ray sources in the SSG CMD region, when $N_{eX}<1$ are expected.
Summing over all observed and expected sources for the globular clusters (bottom line of Table~\ref{obstab})
we predict $N_{eX}<1$ field X-ray source with the appropriate luminosity in our entire globular cluster sample. 
Considering that we observe $N_{oX} = 21$ X-ray sources in the SSG CMD region in globular clusters,
we find a zero percent probability that all the X-ray-detected SSGs in globular clusters are field stars. 

For the open clusters, if we do not consider any kinematic membership information, there is a non-zero 
probability that all SSGs could be field stars.  However, again, the open clusters have the most comprehensive proper-motion and radial-velocity membership information.

In addition to this analysis of the full SSG population, we also estimate a probability that each SSG or RS individually could be a field star (P$_\text{field}$ in Table~\ref{sumtab}), as follows:
\begin{equation} 
\begin{split}
\label{e:field}
P_\mathrm{field} = & \left(1 - P_\mathrm{PM}\right) \times \left(1 - P_\mathrm{RV}\right) \times \\ 
& \begin{cases}
    P_V\left(N_{oV}=1\right) \times P_X\left(N_{oX}=1\right)  ,& \text{if var.\ or X-ray} \\
    P\left(N_o=1\right)   ,& \text{otherwise} \\ 
\end{cases}
\end{split}
\end{equation}

where $P_\mathrm{PM}$ and $P_\mathrm{RV}$ are the fractional kinematic membership probabilities given in Table~\ref{sumtab}.
We assume a 50\% membership probability for stars labeled ``M'', and a 0\% probability for stars labeled ``NM''.
For stars without a $P_\mathrm{PM}$ or $P_\mathrm{RV}$ value in Table~\ref{sumtab}, we simply set the appropriate $P_\mathrm{PM}$ or $P_\mathrm{RV}$
equal to 0 (thereby excluding that membership indicator from the calculation).
$P_V\left(N_{oV}=1\right)$, $P_X\left(N_{oX}=1\right)$ and $P\left(N_o=1\right)$ are calculated following Equation~\ref{e:poisson}, by setting
the observed number, $N_{oV}$, $N_{oX}$ or $N_o$, equal to unity, and using the expected number, $N_{eV}$, $N_{eX}$ or $N_e$, given in Table~\ref{obstab} for the cluster.
We derive expected numbers for the RS stars in a similar manner as described above for the SSGs.
If a given SSG or RS star is either not a photometric variable or X-ray source (with $L_X < 10^{32}$ erg s$^{-1}$), we simply set the respective probability,
$P_V\left(N_{oV}=1\right)$ or $P_X\left(N_{oX}=1\right)$, to unity
(thereby excluding that membership indicator from the calculation).
If the star is neither an X-ray source, with the appropriate luminosity, or a photometric or radial-velocity variable, we replace these probabilities with $P\left(N_o=1\right)$.
Here we assume that these probabilities are independent, and we give each membership indicator equal weight.

These $P_\text{field}$ values provide an estimate of the membership status of each source individually.  From examination of the distribution of these membership values, we select a 
fairly strict cutoff of $P_\text{field} =$~10\%, above which we exclude the source from our member sample used for our following analyses. 
We also exclude the few sources that have $P_\text{field} <$~10\%, but are labeled as ``NM?''.
77\% (43/56) of the SSG and 88\% (7/8) of the RS sources from the literature pass this criteria. 
The remaining have less certain membership status, though may still be cluster members.

In conclusion, even without considering kinematic membership information (as in Table~\ref{obstab}), the probability is remote that all these candidate SSG X-ray sources and photometric 
variables could be field stars (especially in the globular clusters).  We include kinematic information in Equation~\ref{e:field}. The product of these $P_\text{field}$ values, given 
in Table~\ref{sumtab}, provides a more informed estimate of the probability that all sources could be field stars, and is vanishingly small (for all sources, and for the SSG and RS stars 
separately).    
We conclude that it is exceedingly unlikely that all of the SSG sources listed in Table~\ref{sumtab} (X-ray-detected or otherwise) could be field stars, and move forward with 
a sample of highly likely cluster members for our subsequent analyses.

\begin{figure}[!t]
\plotone{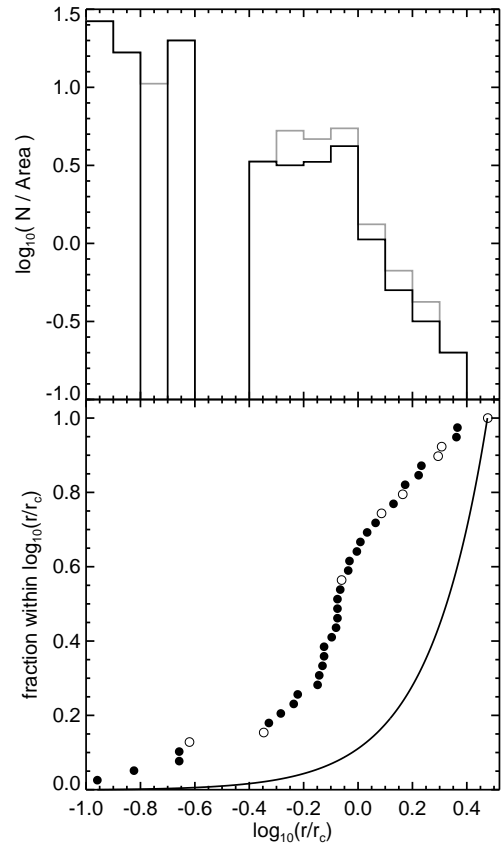}
\caption{
Radial distribution of the SSGs from our sample (i.e., those that fall in the SSG region in at least one color-magnitude combination),
shown in a histogram of surface density (top) as well as the cumulative distribution (bottom). 
In both panels, the black lines and symbols (both open and filled) show SSGs that have the highest likelihood of cluster membership (see Section~\ref{s:agg}), and in 
the top panel we show the additional SSG candidates with the gray lines.
In this figure we exclude SSGs detected at radii beyond the minimum completeness radius of all surveys (of 3.3 core radii in NGC 6121).
In the bottom panel, open/globular cluster sources are plotted with open/filled circles.
Also for comparison in the bottom panel with the solid line, we show the cumulative distribution expected for a population of sources that have a 
uniform surface density as a function of radius, as expected for a field sample.
\label{f:rad}
}
\end{figure}

\subsection{Radial Distributions: Another Indicator of Cluster Membership}
\label{s:rad}

In Figure~\ref{f:rad}, we plot the radial distribution of the SSGs, within our sample of the most likely cluster members.
The specific values plotted in this figure are given in Table~\ref{sumtab}, and are calculated using the cluster centers from \citet{goldsbury:10}.  
Because the surveys become increasingly incomplete at larger radii (see Section~\ref{s:bias}), we limit this analysis to only include 
SSGs found at cluster-centric radii within the minimum completeness radius of all surveys in this study (of 3.3 core radii in NGC 6121).  
Moreover, the survey of SSGs is essentially complete in radius within this limit in each cluster 
(though in some of the globular clusters the surveys were offset from the cluster center, so portions of the clusters may still remain unobserved in X-ray and/or optical).
93\% of the SSGs that have the highest likelihood of cluster membership are found within 3.3 core radii from their respective cluster centers, 
and about 60\% are found within 1 core radius.

For comparison, in the bottom panel of Figure~\ref{sumtab}, we show the cumulative distribution expected for a population of sources distributed uniformly in surface 
density, as would be expected for a field population.  A Kolmogorov-Smirnov (K-S) test shows that we can reject the hypothesis that these SSGs were drawn from a field 
population at very high confidence, with a K-S statistic of 1.7$\times 10^{-6}$.  (We also performed the same test, but excluding the $\omega$~Cen sample, and reach the same
conclusion.)
Thus, this comparison provides further evidence that these sources are indeed 
cluster members.  Furthermore, the full sample of SSGs within the 3.3 core radii completeness limit (and without taking any selection on membership likelihood)
is equally unlikely to be drawn from the field (with a K-S statistic of 1.7$\times 10^{-8}$).  The SSGs are centrally concentrated with respect to their cluster centers, 
as would be expected for any stellar population in a star cluster. 

\section{Aggregate Empirical Characteristics of SSGs}
\label{s:agg}

\begin{figure}[!t]
\plotone{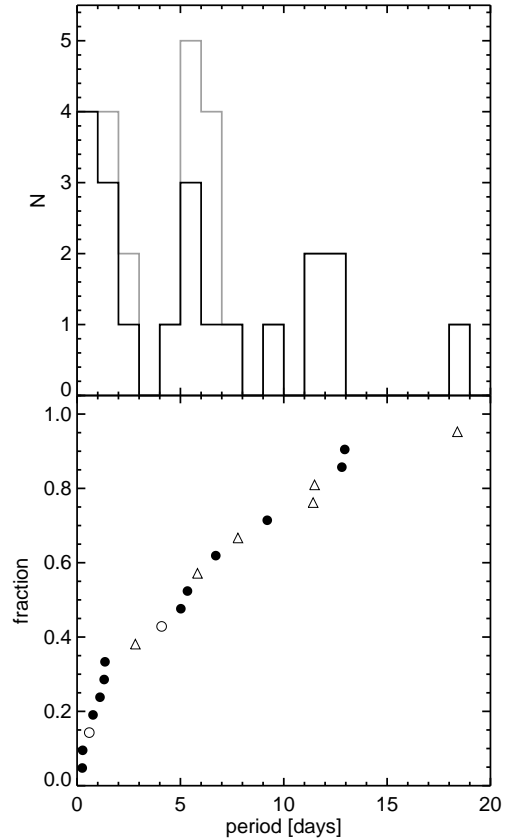}
\caption{
Distribution of the photometric and radial-velocity periods (in days) of the SSGs from our sample (i.e., those that fall in the SSG region in at least one 
color-magnitude combination), shown in histogram (top) and cumulative (bottom) distribution forms.  
In both panels, the black lines and symbols (both open and filled) show SSGs that have the highest likelihood of cluster membership (see Section~\ref{s:agg}, 
and at any cluster-centric radius), and in the top panel we show the additional SSG candidates with the gray lines.
Open/globular cluster sources are plotted with open/filled symbols in the bottom panel.  Here, triangles show periods from radial-velocity 
binary orbital solutions, and circles show periods from photometry.  For sources with both photometric and radial-velocity periods, we plot both. 
\label{f:per}
}
\end{figure}

We list the general properties for this population of SSGs at the beginning of Section~\ref{s:obs}.  Here 
we discuss in further detail the SSG binary properties and period distribution (Figure~\ref{f:per}), 
a comparison of X-ray and optical flux (Figure~\ref{f:xray}), and the frequency of SSGs as a function of host cluster mass (Figure~\ref{f:NvM}).
We will focus here on the most secure cluster members of this SSGs sample, as discussed in Section~\ref{s:contam}

\setlength{\tabcolsep}{2pt}
\begin{deluxetable*}{lccccccccccccc}
\tabletypesize{\tiny}
\tablecaption{Orbital Elements for Sub-subgiant and Red Straggler Radial-Velocity Binaries  \label{bintab}}
\tablehead{\colhead{Cluster} & \colhead{ID} & \colhead{Per$_\text{RV}$} & \colhead{Orbital} & \colhead{$\gamma$} & \colhead{$K$} & \colhead{$e$} & \colhead{$\omega$} & \colhead{$T_0$} & \colhead{$a \sin i$} & \colhead{f($m$)$^\dagger$} &\colhead{$q$} & \colhead{$\sigma$} & \colhead{$N_\text{obs}$}  \\ 
\colhead{} & \colhead{} & \colhead{(days)} & \colhead{Cycles} & \colhead{km s$^{-1}$} & \colhead{km s$^{-1}$} & \colhead{} & \colhead{deg} & \colhead{(HJD$-$2,400,000 d)} & \colhead{$10^6$ km} & \colhead{\Msolar} & \colhead{} & \colhead{km s$^{-1}$} & \colhead{} }
\tablewidth{0pt}
\startdata
NGC 188  & 4289   & 11.4877     & 116.2   & -42.62    & 40.8     & 0.012      & 200       & 50744.5   & 6.44       & 8.07$\times10^{-2}$     & \nodata & 1.11    & 25      \\
         &        & $\pm$0.0009 & \nodata & $\pm$0.23 & $\pm$0.3 & $\pm$0.010 & $\pm$50   & $\pm$1.7  & $\pm$0.05  & $\pm$2.0$\times10^{-3}$ & \nodata & \nodata & \nodata \\
NGC 188  & 3118$^A$ &  11.9022    & 133.8   &  -43.23   & 43.6     & 0.011      & 20        & 50900.3   & 7.14       & 0.656                  &  0.795  &  1.68   &   44 \\
         &          &    $\pm$0.0004 &   \nodata     &      $\pm$0.20 &       $\pm$0.3 &     $\pm$0.006 &        $\pm$30 &       $\pm$1.0 &      $\pm$0.06 &     $\pm$0.012 &     $\pm$0.009 &       &      \\
         & 3118$^B$ &    \nodata    &   \nodata &   \nodata &            54.9 &    \nodata    &   \nodata     &   \nodata     &            8.98 &           0.521 &    \nodata    &  1.98    &   36 \\
         &          &    \nodata    &   \nodata &   \nodata &       $\pm$0.4 &    \nodata    &    \nodata    &    \nodata    &      $\pm$0.07 &     $\pm$0.009 &    \nodata    &  \nodata &  \nodata\\
 
NGC 2682 & 1063   & 18.396      & \nodata & 34.30     & 20.0     & 0.206      & 95       & 47482.19  & 0.0330      & 1.43$\times10^{-2}$     & \nodata & 0.99    & 28      \\
         &        & $\pm$0.005  & \nodata & $\pm$0.20 & $\pm$0.3 & $\pm$0.014 & $\pm$5   & $\pm$0.22 & $\pm$0.0006 & $\pm$7.0$\times10^{-4}$ & \nodata & \nodata & \nodata \\ 
NGC 2682 & 1113$^A$ & 2.823094       & \nodata & 33.4      & 60.6     & 0       & \nodata  & 48916.368  & 0.0157      & 0.544                & 0.703 & 3.15     & 18      \\
         &          & $\pm$0.000014 & \nodata & $\pm$0.4  & $\pm$0.9 & \nodata & \nodata  & $\pm$0.004 & $\pm$0.0003  & $\pm$0.012           & $\pm$0.012 & \nodata & \nodata  \\
         & 1113$^B$ & \nodata       & \nodata & \nodata   & 86.2     & \nodata & \nodata  & \nodata    & 0.0223       & 0.382                & \nodata & 2.10     & 18      \\
         &          & \nodata       & \nodata & \nodata   & $\pm$0.6 & \nodata & \nodata  & \nodata    & $\pm$0.0003  & $\pm$0.012           & \nodata & \nodata & \nodata \\
NGC 6791 & 746    & 11.415      & 33.3    & -48.2     & 21.3     & 0.05       & 320      & 56995.7   & 3.34        & 1.13$\times10^{-2}$     & \nodata & 1.84    & 16      \\
         &        & $\pm$0.007  & \nodata & $\pm$0.8  & $\pm$0.7 & $\pm$0.04 & $\pm$60  & $\pm$2.0  & $\pm$0.11   & $\pm$1.1$\times10^{-3}$ & \nodata & \nodata & \nodata \\
NGC 6791 & 3626   & 5.8248      & 65.3    & -44.7     & 33.2     & 0.013      & 180      & 56997.7   & 2.66        & 2.21$\times10^{-2}$     & \nodata & 1.87    & 16       \\
         &        & $\pm$0.0008 & \nodata & $\pm$0.5  & $\pm$0.8 & $\pm$0.020 & $\pm$100 & $\pm$1.6  & $\pm$0.06   & $\pm$1.5$\times10^{-3}$ & \nodata & \nodata & \nodata \\
NGC 6791 & 15561    & 7.7812      & 48.9    & -45.4     & 27.1     & 0.015      & 0        & 56989.8   & 2.89        & 1.6$\times10^{-2}$      & \nodata & 1.35    & 16      \\
         &        & $\pm$0.0012 & \nodata & $\pm$0.4  & $\pm$0.6 & $\pm$0.019 & $\pm$90  & $\pm$2.0  & $\pm$0.06   & $\pm$1.0$\times10^{-3}$ & \nodata & \nodata & \nodata \\
\enddata
\tablenotetext{}{We provide the orbital period (Per$_\text{RV}$), the center-of-mass radial velocity ($\gamma$), the radial-velocity amplitude ($K$), the eccentricity ($e$), 
 the longitude of periastron ($\omega$), a Julian Date of periastron passage ($T_0$), the projected semi-major axis ($a \sin i$), the mass function$^\dagger$ (f($m$)), the 
mass ratio ($q$, if available), the rms residual velocity from the orbital solution ($\omega$), and the number of RV measurements ($N$).
Orbital parameters for NGC 188 come from \citet{geller:09}, NGC 2682 from \citealt{mathieu:03} and NGC 6791 from \citealt{milliman:16}.
All spectroscopic binaries included here are single-lined, except for NGC 2682 ID 1113, which is double lined; we provide two sets of parameters for 1113 on subsequent rows, 
for the primary and secondary stars, respectively.  \\
$^\dagger$For the two SB2s, NGC 2682 ID 1113 (SSG) and NGC188 ID 3118 (RS), we provide $M_1$sin$^3$($i$) and $M_2$sin$^3$($i$) for the primary ($A$) and secondary ($B$), respectively, in place of the mass function. }
\end{deluxetable*}

\setlength{\tabcolsep}{6pt}

\subsection{Binary Orbital Parameters and Photometric Periods}

The open clusters NGC 188, NGC 2682, NGC 6791 and NGC 6819 are the only clusters with comprehensive and complete multi-epoch radial-velocity 
measurements capable of detecting binary companions and deriving orbital solutions within the relevant regimes in magnitude and orbital period
\citep{geller:08,geller:09,geller:15,hole:09,milliman:14,milliman:16}.
Eight of the eleven SSGs in these clusters (73\%) are detected as binary stars and six have secure kinematic orbital solutions (excluding 3259 in NGC 188, whose orbital 
period is uncertain). We provide the orbital parameters for the six SSGs,
and one RS (NGC 188, ID 3118),
with secure orbital solutions in Table~\ref{bintab}.  
All have orbital periods of 
less than 20 days, with a mean SSG period of about 10 days (and a standard deviation of about 5 days).  
The longest-period binary with an orbital solution, NGC 2682 ID 1063, is also the only eccentric binary 
(with $e = 0.206 \pm 0.014$). \citet{mathieu:03} note that this eccentricity is typical for normal main-sequence and subgiant binaries in M67 with periods of $\sim$20 days, but argues against
a phase of mass transfer or a large evolved primary, as these would both tend to circularize the orbit (in the absence of any sufficiently recent dynamical encounter, 
or a tertiary companion, that could have increased the eccentricity).  

In Figure~\ref{f:per}, we plot the distribution of measured photometric and/or radial-velocity periods for the SSGs with the highest likelihood of cluster membership.
We do not attempt to correct for incompleteness, and therefore this distribution (at least for the photometric periods) may be biased toward shorter periods.
We simply note here that about 90\% (19/21) of the SSGs with measured photometric and/or radial-velocity periods show periodicity at $\lesssim$15 days.

Finally, the SB2, NGC 2682 S1113, is the only SSG that has a dynamically measured mass ratio.  Its mass ratio of 0.7 and the spectroscopic 
temperature and surface gravity measurements of \citet{mathieu:03} point to a main-sequence companion to the SSG.
However, again \citet{mathieu:03} caution that they could not find a self-consistent solution for S1113 that accounts for all of the observations.
The RS star NGC 188 3118 is also an SB2, and has a mass ratio of 0.8.  If 3118 has a mass similar to normal giants in the cluster (1.14 \Msolar), then 
the secondary would have a mass of 0.9 \Msolar\ \citep{geller:09}, and would most likely be a main-sequence star near the turnoff (though further analysis is desirable).

\subsection{X-ray Emission}

\begin{figure*}[!t]
\epsscale{0.95}
\plotone{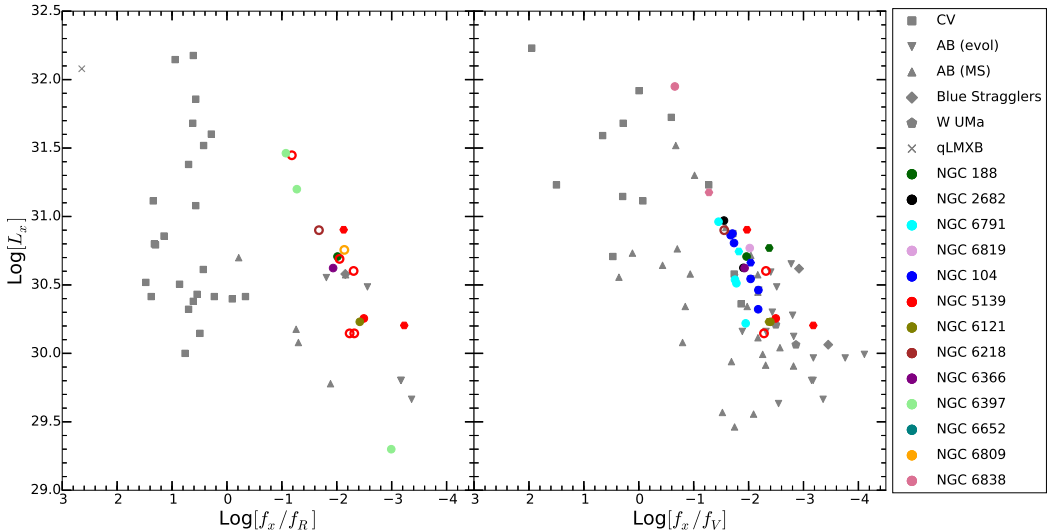}
\epsscale{1.0}
\caption{
Comparison between X-ray luminosities (0.5 - 2.5 keV) and X-ray to optical flux ratios in the $V$ and $R$ optical filters. 
We plot the SSG and RS stars that pass our membership criteria in solid colored circles and hexagons, respectively, and those with uncertain membership in open symbols, 
for all clusters that have measurements in either $V$ or $R$.  Each cluster sample is plotted with a different color.
For comparison, in the gray points, we also plot other X-ray sources that have optical counterparts from the literature, drawing from the same clusters as our SSG sample
(specifically from 47 Tuc, M4, M55, and M67 in the $V$-band and M55, M67, and $\omega$~Cen in the $R$-band).  
The types of sources and related plot symbols for these comparison samples are defined in the figure legend.
For clarity, ``AB (evol)'' are active binaries which appear above the main-sequence turn-off (e.g., RS CVn stars) and ``AB (MS)'' are those which appear below the turnoff
(e.g., BY Dra stars). 
\label{f:xray}
}
\end{figure*}

Next we turn to the X-ray and optical properties of the SSG and RS stars, compared in Figure~\ref{f:xray}.
58\% (25/43) of the SSGs are observed to be X-ray sources, which is likely a lower limit.
There is no obvious preference in CMD locations for SSGs detected in X-ray (see Figure~\ref{f:CMDobs}, where filled symbols mark X-ray sources).

The majority of the cluster X-ray data originate from the \textit{Chandra X-ray Observatory}, because \textit{Chandra's} high spatial resolution enables the 
most accurate associations between X-ray and optical sources, particularly in dense cluster environments. The X-ray luminosity and observation band from the 
literature are quoted in Table~\ref{sumtab}, but for the purpose of comparing clusters, we convert the quantities to a common unabsorbed 2.5--5 keV X-ray band. 
X-ray fluxes are converted with the Portable, Interactive Multi-Mission Simulator (PIMMS)\footnote{PIMMS is maintained by HEASARC and is available here: 
\url{https://heasarc.gsfc.nasa.gov/docs/software/tools/pimms.html}.}, assuming a 1 keV thermal bremsstrahlung spectral model \citep{edmonds:03, heinke:03, haggard:09, cool:13} and the 
nH given in Table~\ref{clustab}.

We plot flux ratios for sources with measurements in the $V$ and $R$ Johnson-Cousins filters, where we have the largest number of 
individual SSG measurements. We account for reddening using the standard relation, with the $(m-M)_V$ and $E(B-V)$ values listed in Table~\ref{clustab}, and 
$A_R/A_V = 0.749$ \citep[from][for reddening by the diffuse interstellar medium]{cardelli:89}. We convert the reddening-corrected magnitudes into 
flux densities with the zero point fluxes from \citet{bessell:79}. The conversion to flux also requires the flux density to be multiplied by the bandwidth of the filter, 
for which we use 415 \AA\ in $R$ and 360 \AA\ in $V$.

In Figure~\ref{f:xray}, we combine the X-ray and optical information for a subset of the SSGs (colored symbols) and several comparison samples (grey symbols) in plots of 
0.5--2.5 keV X-ray luminosity vs.\ X-ray-to-optical flux ratio. The comparison sources are members of the same clusters as our SSG sample
(specifically from 47 Tuc, M4, M55, and M67 in the $V$-band and M55, M67, and $\omega$~Cen in the $R$-band; see references in Section~\ref{s:obs}), and
contain a variety of X-ray-emitting objects, including quiescent low mass X-ray binaries (qLMXB), 
cataclysmic variables (CV), active binaries that are found at or fainter than the MS turnoff (``AB (MS)'', e.g., BY Dra stars), active binaries found beyond the MS turnoff 
(``AB (evol)'', e.g., RS CVn stars), 
blue straggler stars (BSS), and W UMa stars.  We note that the comparison samples are dominated by $\omega$~Cen in $R$ \citep{haggard:13,cool:13} and 47 Tuc in $V$ 
\citep{grindlay:01,edmonds:03}.

The SSGs occupy the same region as active binaries on these X-ray-to-optical diagrams.
Furthermore the comparison sources that occupy the most similar locus are the active binaries that contain evolved stars, i.e., RS CVn binaries.
In both panels, the SSGs follow a track that is separated from most of the BY Dra ABs, CVs, and qLMXBs in the comparison samples.
The RS stars also fall in the same region as the SSGs.
This overlap with the active binaries hints that heightened chromospheric activity may be responsible for the X-ray activity observed for at least some, 
and perhaps many, SSG and RS stars.   The diminished overlap between the SSG and BY Dra stars (which are also chromospherically active) 
is due to the generally lower optical luminosities of most BY Dra's in the comparison samples relative to the SSGs.  The X-ray luminosities of the SSGs 
agree well with both the BY Dra and RS CVn active binary samples.

\subsection{Frequency of Sub-subgiants Across Different Star Clusters}

\begin{figure}[!t]
\epsscale{1.2}
\plotone{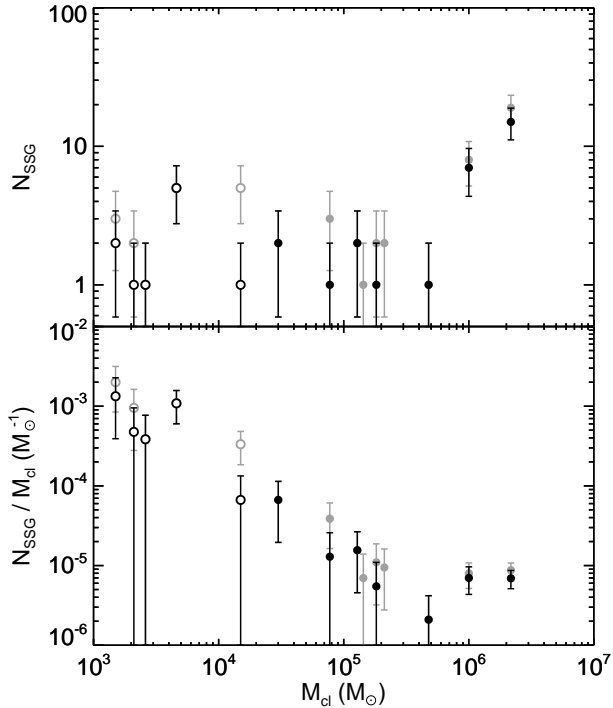}
\caption{
Number (top) and specific frequency (bottom; number of SSGs, $N_\text{SSG}$, divided by the cluster mass, $M_\text{cl}$) of SSGs as a function of the cluster mass for all clusters
with at least one SSG in our sample.  Open/Globular clusters are plotted in open/filled symbols.
Black points include only those SSGs with the highest-likelihood of cluster membership and within the same radial 
completeness limit as shown in Figure~\ref{f:rad}.  Gray points show all SSGs. Error bars show the standard Poisson uncertainties on $N_\text{SSG}$ 
(and we truncate the lower error bars for cases with $N_\text{SSG} = 1$). 
\label{f:NvM}
}
\end{figure}

Figure~\ref{f:NvM} shows the number (top) and  specific frequency (bottom) of SSGs as a function of the cluster mass for all clusters with SSGs in our sample.  
The black points in this figure show only the SSGs that are the most probable cluster members and are within our minimum radius completeness limit, and 
gray points show all SSGs.
The Pearson's correlation coefficient, relating cluster mass and number of SSGs, for the full sample of highly likely cluster members is 0.925.  
However, this value is large because of $\omega$~Cen and 47 Tuc.  Excluding these two clusters (the only clusters in our sample with masses above 10~$^6$\Msolar), 
results in a Pearson's correlation coefficient of -0.278.  Furthermore, a $\chi^2$ test indicates that this limited sample of clusters (still excluding $\omega$~Cen and 47 Tuc)
is consistent with a flat line at the mean value of $\left<N\right>$=1.7 SSGs.  This result holds even if we include $\omega$~Cen and 47 Tuc in this test 
(using $\left<N\right>$=1.7), with a reduced $\chi^2$ value of $\chi^2_\text{red}$=1.91.
We conclude that, for this sample, there is no significant correlation between the number of SSGs and the cluster mass.

This lack of a correlation in $N_\text{SSG}$ vs.\ $M_\text{cl}$ 
results in a clearly visible trend in the specific frequency panel: the number of SSGs per unit mass increases toward lower-mass clusters.  
A $\chi^2$ test shows that the full sample of SSGs is inconsistent with a flat line at the mean value at very high significance ($P(\chi^2) < 10^{-7}$).
The same result holds true for the open and globular clusters individually. 

The open clusters have a significantly higher specific frequency of SSGs than do the globular clusters. 
If we take the mean specific SSG frequency for the open clusters of $\sim10^{-3}$ \Msolar$^{-1}$ and extrapolate that to the globular cluster regime, 
this would predict about 100 SSGs in a globular cluster with a total mass of $10^5$ \Msolar.  
The maximum number of SSGs known in any cluster is 19 in $\omega$~Cen, followed by eight in 47 Tuc 
(both of which have masses $\geq10^6$ \Msolar).
A hint of this trend was also noted by \citet{cool:13} in comparison of their 
$\omega$~Cen SSG sample to those in M67 and NGC 6791.   This is now verified with the larger sample of SSGs studied here, and we discuss the 
implications of this result in Section~\ref{s:discuss}.

\subsection{Summary of Demographics}

We conclude that the SSG region has an 
elevated frequency of variable stars and X-ray sources relative to stars of similar luminosities in the normal regions of a CMD.  
The X-ray and optical luminosities of these sources are most similar to those of active binaries, and specifically RS CVn stars.
Likewise, at least 14 of the SSGs in our sample are H$\alpha$ emitters (and not all have been observed for H$\alpha$ emission), another common characteristic of RS CVn stars.
Importantly, the open clusters NGC 188, NGC 2682, NGC 6819 and NGC 6791 have comprehensive three-dimensional membership analyses;
these long-term multi-epoch radial-velocity surveys reveal a 73\% (8/11) binary frequency.
Thus, at least for the open clusters, we conclude that the SSGs have a high frequency of binaries relative to the normal stars (which for these clusters 
typically have measured spectroscopic binaries frequencies, with the same completeness limits in orbital periods, closer to $\sim$25\%).
Taking the entire sample of SSG members, we find 65\% (28/43) to be photometric and/or radial-velocity variables, 
21 of which are radial-velocity binaries.
Of perhaps equal importance to the binary and X-ray properties is the trend of increasing specific frequency of SSGs toward lower mass clusters. 
Apparently open clusters are more efficient, per unit mass, at producing SSGs than globular clusters.

\section{Discussion and Conclusions}
\label{s:discuss}

This is the first paper in a series studying the origin of the sub-subgiant (SSG) stars, found
redward of the MS and fainter than the normal giant branch (see Figure~\ref{f:CMDobs}).  
Here we identify from the literature a sample of 56 SSGs 
(plus one candidate SSG discussed in the literature but without reliable color information due to short timescale variability), 
and 8 red stragglers (RSs) in 16 star clusters, 
including both open and globular clusters (Table~\ref{sumtab}).
We identify 43 SSGs from this sample with the highest likelihood of cluster membership.
This sample has the following important empirical characteristics:
\begin{enumerate}
\item They occupy a unique location on a CMD, redward of the normal MS stars but fainter than the subgiant branch, where normal single-star evolution does not predict stars.
\item $\geq$58\% (25/43) of the SSGs are observed to be X-ray sources, with typical luminosities of order 10$^{30-31}$ erg s$^{-1}$, consistent with active binaries.
\item $\geq$33\% (14/43) of the SSGs are known to exhibit H$\alpha$ emission (including $>$50\% in the MUSE sample).
\item $\geq$65\% (28/43) of the SSGs are known photometric and/or radial-velocity variables, with typical periods of $\lesssim$15 days.
\item $\geq$75\% (21/28) of the variable SSGs are radial-velocity binaries.
\item The specific frequency of SSGs increases toward lower-mass star clusters.
\end{enumerate}
All of the percentages given above are lower limits (more complete demographics require further observations).

Most of these sources have kinematic membership data indicative of cluster membership.  
In Sections~\ref{s:contam}~and~\ref{s:agg} we also argue that the X-ray luminosities, photometric 
variability and their centrally concentrated radial distribution with respect to their cluster centers (Figure~\ref{f:rad}), provides additional strong evidence that these 
sources are indeed cluster members.

The large fraction of short-period variables (most either inferred or confirmed to be due to binary companions), and 
the similarities in the X-ray characteristics between the SSGs and active binaries (Figure~\ref{f:xray}), suggest that the SSGs form through binary-mediated channels.
The \textit{Kepler} ``No-Man's-Land'' stars (Section~\ref{ss:kepler}), if confirmed as SSGs, also indicate that SSGs can form in isolation through 
binary evolution channels.
Furthermore, the X-ray and optical luminosities and the H$\alpha$ emission seen in many of these SSGs are similar to RS CVn binaries.

It is well known that dynamics can both create and destroy binaries. 
We find that the specific frequency of SSGs increases toward smaller cluster masses (Figure~\ref{f:NvM}), where dynamical encounters are on average 
less frequent and less energetic.  Thus this trend may indicate that dynamics inhibits the formation of SSGs, perhaps by disrupting or modifying the binary progenitors to SSGs. 
Toward the highest mass clusters there is a hint at a flattening to this distribution in specific SSG frequency, which may 
be indicative of an increased efficiency of dynamical production.  

An important next step is to search for additional SSGs, particularly in the globular clusters, that may be revealed from comprehensive proper-motion (and radial-velocity)
membership analyses.  Still a typical globular cluster would need $\sim$100 SSGs to be consistent with the mean specific frequency of SSGs in the open clusters, 
which seems exceedingly unlikely for the globular clusters studied here.

Of additional great interest are the wide variety of evolutionary states of the companions known for these SSGs.  Within our SSG sample, there is one MSP (and possibly a second) in NGC 6397, a 
candidate black hole companion in NGC 6121 (M4), a massive compact companion (either neutron star or black hole) in NGC 6652, a MS companion near the turnoff in NGC 2682 (M67), 
and three SSGs in W UMa binaries (in NGC 188, $\omega$~Cen and NGC 6397).  
This is far from a complete list, but in other stellar systems that do not follow standard stellar evolution theory, 
the companions provide significant insight into the most active formation channels.  
(For example, the white dwarf companions to blue stragglers in NGC 188 found by \citealt{geller:11} and \citealt{gosnell:15} point directly to a past stage of mass transfer as 
their formation mechanism.) 
The compact object companions to these SSGs are particularly intriguing. To obtain an orbital period on the order of 10 days in a binary containing a neutron star 
or black hole, presumably either the system formed early and went through a previous stage of mass transfer or common envelope to shrink the orbit, or the system was formed
more recently by a dynamical encounter, perhaps involving a tidal-capture scenario.  
Regardless of the specific evolutionary histories, the diversity of companions amongst the SSGs may offer important guidance in developing theories of SSG formation, 
which we discuss in subsequent papers.

Further observations aimed at determining the companion stars to these SSGs (and RSs) would be very valuable.
Also, recall that not all of these sources have the relevant observations to determine an X-ray luminosity or variability period.
Therefore it is possible that even more of these SSGs are indeed X-ray sources and/or variables with similar luminosities and periods, respectively, to those shown 
in Figures~\ref{f:per} and~\ref{f:xray}.  Follow-up observations of these sources are desirable.
Additionally, the nearly 10,000 stars in the ``No-Man's-Land'' from \textit{Kepler} \citep{batalha:13,huber:14}, are a very interesting population for follow-up work.
These may be field SSGs (that may have formed without the need for cluster dynamics), and we will investigate these ``No-Man's-Land'' stars in more detail in a future paper.  

Though we focus in the majority of this paper on the SSG stars, we return here briefly to comment on the RS stars.  There are eight total RS stars in this sample 
(in NGC 188, NGC 6791, NGC 7142, $\omega$~Cen and NGC 6752), seven of which would pass our criteria for high likelihood of cluster membership 
(and one of these is outside of the radial completeness that we have set above).  Thus the number of sources 
is too small to perform any rigorous statistical tests.  Of these seven RS stars that would pass our membership criteria, four are X-ray sources, and three are photometric 
or radial-velocity variables.
Thus, apart from their somewhat different location on the CMD, the RS stars appear to have similar characteristics as the SSGs.  
As we will discuss in future papers, certain formation channels predict that the SSG and RS stars are related through evolution, where one may be 
the progenitor of the other.  

Previous authors have suggested various dynamical and interacting-binaries scenarios to explain the SSGs, though none present detailed models of SSG formation (or evolution).  
For instance, \citet{albrow:01} posit, with regards to the sources in 47 Tuc and M67 (the only SSGs known at the time), that  
``a plausible explanation for these stars is a deflated radius from subgiant or giant origins as the result of 
mass transfer initiated by Roche lobe contact by the evolved star, for which the secondary has a lower mass'', where PC-V11 has a WD companion and 
the others have MS companions. They liken these stars to BV Centauri, which \citet{gilliland:82} suggests is a long-period CV variable with 
a subgiant donor, and an expected lifetime that may exceed $10^9$ years.  
\citet{hurley:05} briefly discuss the formation of a SSG star in their $N$-body model of M67 through a common-envelope merger event
(created within the parameterized Binary Stellar Evolution code BSE, \citealt{hurley:02}).
Empirically, there is evidence in some systems for past, and possibly ongoing mass transfer (e.g., the MSP system in NGC 6397). 
We also note that some of the main-sequence -- main-sequence collision models of \citet{sills:05} evolve through the SSG region.  

We perform an in-depth investigation of specific SSG formation routes in Leiner et al.\ (in prep).
In this and subsequent papers we will investigate two general modes of SSG formation, (i) through 
isolated binary evolution and (ii) through dynamical processes (also likely involving binaries). These papers will provide the first detailed models of SSG formation 
and will investigate the relative formation frequencies that each model predicts.  The observations and demographic information provided in this paper solidify 
SSGs as a new class of stars that defies standard single-star evolution theory.
These stars may prove to be very important test cases for both binary evolution and star cluster dynamics modeling.

\acknowledgments
A.M.G.\ acknowledges support from HST grant AR-13910 and a National Science Foundation Astronomy and Astrophysics Postdoctoral Fellowship Award No.\ AST-1302765.
E.M.L.\ was supported by a Grant-In-Aid of Research from Sigma Xi, the Scientific Research Society.
A.B.\ acknowledges support from HST grant AR-12845.
S.K.\ received funding through BMBF Verbundforschung (project MUSE-AO, grant 05A14MGA).
A.S.\ is supported by the Natural Sciences and Engineering Research Council of Canada.

\clearpage
\LongTables
\begin{landscape}

\setlength{\tabcolsep}{3pt}
\begin{deluxetable}{lllllccccccccccccccccc}
\tabletypesize{\tiny}
\tablecaption{Sub-subgiant Summary Table \label{sumtab}}
\tablehead{\colhead{Cluster} & \colhead{Other} & \colhead{ID} & \colhead{RA} & \colhead{Dec} & \colhead{P$_{\mathrm{PM}}$} & \colhead{P$_{\mathrm{RV}}$} & \colhead{P$_{\mathrm{field}}$} & \colhead{$r/r_c$} & \colhead{$U$} & \colhead{$B$} & \colhead{$V$} & \colhead{$R$} & \colhead{$I$} & \colhead{$J$} & \colhead{$H$} & \colhead{$K$} & \colhead{$L_X$} & \colhead{$X$-band} & \colhead{Per$_\text{RV}$} & \colhead{Per$_\text{phot}$} & \colhead{SSG/RS/N} \\ 
\colhead{} & \colhead{} & \colhead{} & \colhead{(J2000.0)} & \colhead{(J2000.0)} & \colhead{[\%]} & \colhead{[\%]} & \colhead{[\%]} & \colhead{} & \colhead{} & \colhead{} & \colhead{} & \colhead{} & \colhead{} & \colhead{} & \colhead{} & \colhead{} & \colhead{[erg s$^{-1}$]} & \colhead{[keV]} & \colhead{[d]} & \colhead{[d]} & \colhead{} } 
\tablewidth{0pt}
\startdata
&&&&&&&&&&&&&&&&&&&\\
\multicolumn{21}{l}{\textbf{Open Clusters}} \\
&&&&&&&&&&&&&&&&&&&\\
                  NGC 188 &      \nodata &       1141 &          00:44:44.50 &          +85:32:16.4 &      88 &      98 &     0.002 &  4.34 &    \nodata &      15.28 &      14.11 &    \nodata &    \nodata &       12.1 &      11.54 &      11.42 &      1.02$\times 10^{31}$ &   0.1-2.4  &    \nodata &    \nodata & 0/1/0 \\
                  NGC 188 &      \nodata &       3118 &          00:32:21.20 &          +85:18:38.9 &      34 &      88 &     0.164 &  4.59 &    \nodata &      15.78 &      14.65 &    \nodata &    \nodata &      12.51 &      11.95 &      11.73 &                   \nodata &    \nodata &       11.9 &    \nodata & 0/1/0 \\
                  NGC 188 &      \nodata &       3259 &          00:34:59.19 &          +85:18:42.8 &      71 &       M &     3.213 &  3.82 &    \nodata &      17.10 &      16.12 &    \nodata &    \nodata &      14.29 &      13.99 &      13.73 &                   \nodata &    \nodata &      102?  &    \nodata & 1/0/0 \\
                  NGC 188 &      \nodata &       4289 &          00:42:40.39 &          +85:16:49.5 &      98 &      98 &     0.001 &  1.46 &      16.94 &      16.24 &      15.30 &      14.77 &      14.20 &      13.54 &      13.00 &      12.88 &      4.70$\times 10^{30}$ &   0.5-2.0  &      11.49 &    \nodata & 10/0/0 \\
                  NGC 188 &      \nodata &       4989 &          00:48:22.65 &          +85:15:55.3 &      95 & \nodata &     1.108 &  0.45 &      17.50 &      17.09 &      16.14 &      15.57 &      15.06 &      14.44 &      13.84 &      13.83 &                   \nodata &    \nodata &    \nodata &       0.59 & 9/0/1 \\
                 NGC 2158 &      \nodata &        V44 &           06:07:05.8 &          +24:08:51.4 &   44/80 & \nodata &    16.944 &  0.93 &    \nodata &    \nodata &      17.41 &      16.70 &    \nodata &    \nodata &    \nodata &    \nodata &                   \nodata &    \nodata &    \nodata &        2.8 & 1/0/0 \\
                 NGC 2158 &      \nodata &        V48 &           06:07:06.2 &          +24:02:10.3 &   41/69 & \nodata &    26.264 &  1.24 &    \nodata &    \nodata &      17.11 &      16.41 &    \nodata &    \nodata &    \nodata &    \nodata &                   \nodata &    \nodata &    \nodata &       6.34 & 1/0/0 \\
                 NGC 2158 &      \nodata &        V49 &           06:07:10.2 &          +24:10:19.5 &  26/51  & \nodata &    41.514 &  1.45 &    \nodata &    \nodata &      17.17 &      16.48 &    \nodata &    \nodata &    \nodata &    \nodata &                   \nodata &    \nodata &    \nodata &       6.73 & 1/0/0 \\
                 NGC 2158 &      \nodata &        V90 &           06:07:36.2 &          +24:05:25.4 &      94 & \nodata &     5.083 &  2.32 &    \nodata &    \nodata &      16.58 &      15.92 &    \nodata &    \nodata &    \nodata &    \nodata &                   \nodata &    \nodata &    \nodata &      12.81 & 1/0/0 \\
                 NGC 2158 &      \nodata &        V95 &           06:07:18.8 &          +24:01:40.1 &  53/70  & \nodata &    25.417 &  1.74 &    \nodata &    \nodata &      16.70 &      15.92 &    \nodata &    \nodata &    \nodata &    \nodata &                   \nodata &    \nodata &    \nodata &        var & 1/0/0 \\
                 NGC 2682 &          M67 &      S1063 &          08:51:13.37 &          +11:51:40.2 &      97 &      98 &     0.013 &  0.87 &      15.56 &      14.84 &      13.79 &    \nodata &      12.59 &      11.66 &      11.06 &      10.96 &      1.30$\times 10^{31}$ &     0.3-7  &       18.4 &    \nodata & 6/0/0 \\
                 NGC 2682 &          M67 &      S1113 &           08:51:25.4 &          +12:02:56.5 &      98 &      98 &     0.008 &  3.36 &       15.3 &      14.78 &      13.77 &    \nodata &    \nodata &      11.67 &      11.12 &      10.97 &       7.3$\times 10^{30}$ &   0.1-2.4  &       2.82 &    \nodata & 3/0/0 \\
                 NGC 6791 &      \nodata &         83 &          19:20:10.61 &          +37:51:11.2 &      96 &      85 &     0.600 &  3.00 &    \nodata &    \nodata &       18.3 &    \nodata &       16.8 &      15.79 &      15.04 &      14.78 &                   \nodata &    \nodata &    \nodata &    \nodata & 1/0/0 \\
                 NGC 6791 &      \nodata &        746 &          19:20:21.48 &          +37:48:21.6 &      99 &      95 &     0.044 &  2.03 &    \nodata &      19.31 &      17.96 &    \nodata &      16.56 &      15.53 &      14.81 &      14.71 &      4.50$\times 10^{30}$ &     0.3-7  &     11.419 &      13.83 & 3/0/0 \\
                 NGC 6791 &      \nodata &       3626 &          19:20:38.88 &          +37:49:04.3 &      99 &      63 &     0.323 &  1.22 &    \nodata &      19.12 &      17.96 &    \nodata &      16.67 &      15.80 &      15.25 &      15.06 &      4.80$\times 10^{30}$ &     0.3-7  &     5.8251 &       6.37 & 3/0/0 \\
                 NGC 6791 &      \nodata &       6371 &          19:20:47.88 &          +37:46:37.2 &      99 &      M? &     0.083 &  0.33 &    \nodata &      18.42 &      17.27 &    \nodata &      15.86 &      14.89 &      14.30 &      14.04 &      7.70$\times 10^{30}$ &     0.3-7  &        var &       3.19 & 0/3/0 \\
                 NGC 6791 &      \nodata &       7011 &          19:20:49.86 &          +37:45:50.7 &      99 &      M? &     0.436 &  0.24 &    \nodata &      19.33 &      18.27 &    \nodata &      17.15 &      16.43 &      15.87 &      15.24 &      2.30$\times 10^{30}$ &     0.3-7  &        var &       4.09 & 3/0/0 \\
                 NGC 6791 &      \nodata &      15561 &          19:21:25.22 &          +37:45:49.8 &      99 &      84 &     0.139 &  1.97 &    \nodata &      18.99 &      17.65 &    \nodata &      16.13 &      15.20 &      14.50 &      14.49 &      1.27$\times 10^{31}$ &     0.3-7  &     7.7815 &       7.64 & 1/2/0 \\
                 NGC 6819 &      \nodata &      52004 &          19:41:25.89 &          +40:12:23.6 &      99 & \nodata &     0.911 &  0.75 &    \nodata &      16.49 &      15.65 &    \nodata &    \nodata &      14.14 &      13.75 &      13.76 &      9.30$\times 10^{30}$ &  0.2-10.0  &    \nodata &    \nodata & 1/0/0 \\
                 NGC 7142 &      \nodata &         V4 &          21:44:55.98 &          +65:45:50.0 &      91 & \nodata &     7.435 &  0.44 &    \nodata &      17.32 &      15.43 &    \nodata &      13.37 &      11.97 &      11.00 &      10.77 &                   \nodata &    \nodata &    \nodata &        var & 0/3/0 \\
&&&&&&&&&&&&&&&&&&&&\\
\multicolumn{21}{l}{\textbf{Globular Clusters}} \\
&&&&&&&&&&&&&&&&&&&\\
                  NGC 104 &       47 Tuc &    PC1-V11 &          00:24:04.65 &          -72:04:55.3 & \nodata & \nodata &  $<$0.001 &  0.15 &      17.83 &    \nodata &      17.37 &    \nodata &      16.41 &    \nodata &    \nodata &    \nodata &      7.50$\times 10^{30}$ &   0.5-2.5  &    \nodata &      1.108 & 2/0/1 \\
                  NGC 104 &       47 Tuc &    PC1-V41 &          00:24:08.76 &          -72:05:07.7 &       M & \nodata &  $<$0.001 &  0.55 &      18.54 &    \nodata &      17.61 &    \nodata &      16.89 &    \nodata &    \nodata &    \nodata &      2.10$\times 10^{30}$ &   0.5-2.5  &    \nodata &       5.02 & 2/0/1 \\
                  NGC 104 &       47 Tuc &    PC1-V48 &          00:24:10.46 &          -72:05:00.0 &       M &      M? &  $<$0.001 &  0.62 &      18.54 &    \nodata &      17.37 &    \nodata &      16.35 &    \nodata &    \nodata &    \nodata &      3.50$\times 10^{30}$ &   0.5-2.5  &        var &       6.71 & 3/0/0 \\
                  NGC 104 &       47 Tuc &    WF2-V31 &          00:24:15.16 &          -72:04:43.6 &       M &       M &  $<$0.001 &  1.19 &      18.87 &    \nodata &      17.56 &    \nodata &      16.39 &    \nodata &    \nodata &    \nodata &      7.30$\times 10^{30}$ &   0.5-2.5  &        var &       5.34 & 3/0/0 \\
                  NGC 104 &       47 Tuc &    WF2-V32 &          00:24:13.76 &          -72:03:34.6 &       M &      M? &  $<$0.001 &  2.31 &      18.18 &    \nodata &      17.12 &    \nodata &      16.10 &    \nodata &    \nodata &    \nodata &                   \nodata &    \nodata &        var &        9.2 & 1/2/0 \\
                  NGC 104 &       47 Tuc &    WF4-V18 &          00:24:12.01 &          -72:05:12.5 &     NM? &       M &  $<$0.001 &  0.94 &      18.11 &    \nodata &      17.26 &    \nodata &      16.37 &    \nodata &    \nodata &    \nodata &                   \nodata &    \nodata &        var &        5.9 & 3/0/0 \\
                  NGC 104 &       47 Tuc &         W4 &          00:24:13.20 &          -72:04:51.1 & \nodata & \nodata &  $<$0.001 &  0.92 &      18.20 &    \nodata &      17.48 &    \nodata &      16.63 &    \nodata &    \nodata &    \nodata &       6.4$\times 10^{30}$ &   0.5-2.5  &    \nodata &    \nodata & 3/0/0 \\
                  NGC 104 &       47 Tuc &        W38 &          00:24:04.64 &          -72:04:46.0 & \nodata & \nodata &  $<$0.001 &  0.22 &      18.04 &    \nodata &      17.23 &    \nodata &      16.35 &    \nodata &    \nodata &    \nodata &      2.90$\times 10^{30}$ &   0.5-2.5  &    \nodata &    \nodata & 3/0/0 \\
                  NGC 104 &       47 Tuc &        W43 &          00:24:04.20 &          -72:04:43.6 & \nodata & \nodata &  $<$0.001 &  0.31 &      18.46 &    \nodata &      17.08 &    \nodata &    \nodata &    \nodata &    \nodata &    \nodata &      4.60$\times 10^{30}$ &   0.5-2.5  &    \nodata &    \nodata & 0/1/0 \\
                 NGC 5139 & $\omega$ Cen &         12 &          13:26:58.36 &          -47:32:46.7 &      99 &       M &     0.500 &  0.93 &      19.19 &      18.62 &      17.64 &      17.18 &      16.71 &    \nodata &    \nodata &    \nodata &                   \nodata &    \nodata &    \nodata &    \nodata & 10/0/0 \\
                 NGC 5139 & $\omega$ Cen &         16 &          13:26:34.20 &          -47:32:14.8 &      93 &       M &     1.549 &  0.86 &      20.02 &      19.82 &      18.76 &      18.58 &      18.09 &    \nodata &    \nodata &    \nodata &                   \nodata &    \nodata &        var &    \nodata & 8/0/2 \\
                 NGC 5139 & $\omega$ Cen &         22 &          13:27:04.24 &          -47:31:33.9 &      96 &       M &     2.000 &  0.84 &      19.47 &      19.83 &      18.83 &      18.52 &      18.11 &    \nodata &    \nodata &    \nodata &                   \nodata &    \nodata &    \nodata &    \nodata & 4/0/6 \\
                 NGC 5139 & $\omega$ Cen &         23 &          13:26:08.26 &          -47:31:26.7 &      84 &       M &     3.541 &  1.49 &      19.53 &      19.14 &      18.59 &      18.34 &      17.85 &    \nodata &    \nodata &    \nodata &                   \nodata &    \nodata &    \nodata &       0.25 & 5/0/5 \\
                 NGC 5139 & $\omega$ Cen &         28 &          13:27:06.84 &          -47:30:58.7 &      95 &       M &     1.107 &  0.83 &      18.54 &      19.08 &      17.98 &    \nodata &      17.18 &    \nodata &    \nodata &    \nodata &                   \nodata &    \nodata &        var &    \nodata & 5/0/1 \\
                 NGC 5139 & $\omega$ Cen &         35 &          13:27:08.25 &          -47:30:14.3 &      96 &       M &     2.000 &  0.80 &      18.36 &      18.63 &      17.63 &      17.29 &      16.81 &    \nodata &    \nodata &    \nodata &                   \nodata &    \nodata &    \nodata &    \nodata & 8/0/2 \\
                 NGC 5139 & $\omega$ Cen &         36 &          13:27:05.83 &          -47:30:02.7 &      96 &       M &     0.885 &  0.71 &      18.17 &      18.23 &      17.23 &      16.85 &      16.21 &    \nodata &    \nodata &    \nodata &                   \nodata &    \nodata &        var &    \nodata & 9/0/1 \\
                 NGC 5139 & $\omega$ Cen &         37 &          13:26:31.57 &          -47:29:58.0 &     100 &       M &  $<$0.001 &  0.61 &      18.55 &      17.42 &      16.38 &      15.92 &      15.39 &    \nodata &    \nodata &    \nodata &                   \nodata &    \nodata &    \nodata &    \nodata & 0/8/2 \\
                 NGC 5139 & $\omega$ Cen &         42 &          13:26:23.72 &          -47:29:20.1 &      99 &       M &     0.221 &  0.84 &      19.35 &      18.75 &      17.78 &      17.38 &      16.84 &    \nodata &    \nodata &    \nodata &                   \nodata &    \nodata &        var &    \nodata & 10/0/0 \\
                 NGC 5139 & $\omega$ Cen &         49 &          13:26:38.35 &          -47:25:41.2 &      91 &       M &     1.992 &  0.72 &      19.27 &      18.75 &      17.73 &      17.34 &      16.68 &    \nodata &    \nodata &    \nodata &                   \nodata &    \nodata &        var &    \nodata & 10/0/0 \\
                 NGC 5139 & $\omega$ Cen &         50 &          13:27:08.30 &          -47:25:41.3 &      94 &       M &     1.328 &  0.99 &      19.47 &      19.55 &      18.55 &      18.29 &      17.72 &    \nodata &    \nodata &    \nodata &                   \nodata &    \nodata &        var &    \nodata & 8/0/2 \\
                 NGC 5139 & $\omega$ Cen &         58 &          13:26:30.68 &          -47:24:25.8 &      99 &       M &     0.221 &  1.08 &      19.59 &      19.06 &      18.08 &      17.58 &      17.03 &    \nodata &    \nodata &    \nodata &                   \nodata &    \nodata &        var &    \nodata & 10/0/0 \\
                 NGC 5139 & $\omega$ Cen &         68 &          13:26:16.57 &          -47:22:43.5 &      94 &       M &     3.000 &  1.67 &      19.45 &      19.44 &      18.33 &      17.53 &      16.64 &    \nodata &    \nodata &    \nodata &                   \nodata &    \nodata &    \nodata &    \nodata & 10/0/0 \\
                 NGC 5139 & $\omega$ Cen &        13b &          13:26:50.53 &          -47:29:18.1 & \nodata &       M &    10.321 &  0.16 &    \nodata &      18.27 &    \nodata &      17.14 &    \nodata &    \nodata &    \nodata &    \nodata &      4.90$\times 10^{30}$ &   0.5-2.5  &    \nodata &    \nodata & 1/0/0 \\
                 NGC 5139 & $\omega$ Cen &        22e &          13:26:59.93 &          -47:28:09.7 &      94 & \nodata &     1.239 &  0.47 &      19.37 &      17.97 &      17.23 &      16.42 &      15.97 &    \nodata &    \nodata &    \nodata &      8.00$\times 10^{30}$ &   0.5-2.5  &    \nodata &    \nodata & 3/6/1 \\
                 NGC 5139 & $\omega$ Cen &        24f &          13:26:37.29 &          -47:29:42.9 &      95 & \nodata &     0.041 &  0.41 &      17.73 &      16.97 &      15.96 &      15.41 &      14.93 &    \nodata &    \nodata &    \nodata &      1.60$\times 10^{30}$ &   0.5-2.5  &    \nodata &    \nodata & 0/9/1 \\
                 NGC 5139 & $\omega$ Cen &        32f &          13:27:05.33 &          -47:28:08.8 &      43 & \nodata &    11.766 &  0.65 &      18.02 &      18.27 &      17.12 &      16.72 &      15.97 &    \nodata &    \nodata &    \nodata &      4.00$\times 10^{30}$ &   0.5-2.5  &    \nodata &    \nodata & 4/4/2 \\
                 NGC 5139 & $\omega$ Cen &        34b &          13:26:37.44 &          -47:30:53.3 & \nodata & \nodata &    20.643 &  0.56 &    \nodata &      18.77 &    \nodata &      17.42 &    \nodata &    \nodata &    \nodata &    \nodata &      2.80$\times 10^{31}$ &   0.5-2.5  &    \nodata &    \nodata & 1/0/0 \\
                 NGC 5139 & $\omega$ Cen &        41h &          13:26:43.96 &          -47:24:42.6 & \nodata & \nodata &    20.643 &  0.86 &    \nodata &      19.17 &    \nodata &      18.04 &    \nodata &    \nodata &    \nodata &    \nodata &      1.40$\times 10^{30}$ &   0.5-2.5  &    \nodata &    \nodata & 1/0/0 \\
                 NGC 5139 & $\omega$ Cen &        42c &          13:27:09.65 &          -47:27:28.9 &      75 & \nodata &     5.161 &  0.84 &      19.03 &      19.07 &      18.35 &      17.83 &      17.46 &    \nodata &    \nodata &    \nodata &      1.40$\times 10^{30}$ &   0.5-2.5  &    \nodata &    \nodata & 9/0/1 \\
                 NGC 5139 & $\omega$ Cen &        43c &          13:27:06.91 &          -47:30:09.4 &      79 & \nodata &     4.335 &  0.75 &      18.65 &      18.47 &      17.54 &      17.12 &      16.46 &    \nodata &    \nodata &    \nodata &      1.80$\times 10^{30}$ &   0.5-2.5  &    \nodata &    \nodata & 10/0/0 \\
                 NGC 6121 &           M4 &        CX8 &          16:23:31.48 &          -26:30:57.8 &       M & \nodata &  $<$0.001 &  1.02 &      17.84 &      17.67 &       16.7 &      16.03 &      15.39 &      14.57 &      13.99 &      13.79 &      1.70$\times 10^{30}$ &    0.5-2.5 &    \nodata &       0.77 & 10/0/0 \\
                 NGC 6121 &           M4 &       CX10 &          16:23:35.05 &          -26:31:19.7 &       M &       M &  $<$0.001 &  0.22 &    \nodata &    \nodata &      16.77 &    \nodata &      15.25 &    \nodata &    \nodata &    \nodata &      1.70$\times 10^{30}$ &   0.5-2.5  &    \nodata &    \nodata & 1/0/0 \\
                 NGC 6218 &          M12 &       CX2b &          16:47:18.40 &          -01:56:53.6 &     NM? & \nodata &     3.249 &  0.71 &    \nodata &      19.04 &      18.32 &      17.54 &    \nodata &    \nodata &    \nodata &    \nodata &      1.10$\times 10^{31}$ &     0.3-7  &    \nodata &    \nodata & 3/0/0 \\
                 NGC 6366 &      \nodata &        CX5 &          17:27:44.14 &          -05:04:27.2 & \nodata & \nodata &     1.638 &  0.11 &    \nodata &      20.67 &      19.01 &      18.11 &      17.05 &    \nodata &    \nodata &    \nodata &      4.20$\times 10^{30}$ &   0.5-2.5  &    \nodata &    \nodata & 5/0/1 \\
                 NGC 6397 &      \nodata &        U12 &          17:40:44.62 &          -53:40:41.5 &       M &       M &     0.004 &  9.83 &    \nodata &      17.40 &    \nodata &      16.18 &    \nodata &    \nodata &    \nodata &    \nodata &      2.20$\times 10^{31}$ &     0.3-8  &        var &       1.35 & 1/0/0 \\
                 NGC 6397 &      \nodata &        U18 &          17:40:42.61 &          -53:40:27.5 &       M &       M &     0.004 &  1.72 &    \nodata &      17.48 &    \nodata &      16.02 &    \nodata &    \nodata &    \nodata &    \nodata &      3.20$\times 10^{31}$ &     0.5-6  &        var &       1.31 & 1/0/0 \\
                 NGC 6397 &      \nodata &        U92 &          17:40:43.92 &          -53:40:35.2 &       M &       M &     0.004 &  6.68 &    \nodata &      18.23 &    \nodata &      16.63 &    \nodata &    \nodata &    \nodata &    \nodata &      2.20$\times 10^{29}$ &     0.5-6  &        var &       0.27 & 1/0/0 \\
                 NGC 6652 &      \nodata &          B &          18:35:43.65 &          -32:59:26.8 & \nodata & \nodata &   \nodata &  1.07 &    \nodata &    \nodata &         20 &    \nodata &    \nodata &    \nodata &    \nodata &    \nodata &      2.20$\times 10^{36}$ &      2-10  &    \nodata &        var & 0/0/0 \\
                 NGC 6752 &      \nodata &        V19 &          19:11:13.04 &          -59:55:17.6 &     NM? & \nodata &    46.541 & 17.42 &    \nodata &      17.39 &      16.32 &    \nodata &    \nodata &    \nodata &    \nodata &    \nodata &                   \nodata &    \nodata &    \nodata &       6.18 & 1/0/0 \\
                 NGC 6752 &      \nodata &        V20 &          19:10:50.50 &          -59:57:37.2 &     NM? & \nodata &    86.466 &  5.56 &    \nodata &      16.83 &      16.09 &    \nodata &    \nodata &    \nodata &    \nodata &    \nodata &                   \nodata &    \nodata &    \nodata &      $>$ 8 & 0/1/0 \\
                 NGC 6752 &      \nodata &        V22 &          19:10:29.67 &          -59:56:24.1 & \nodata & \nodata &    46.541 & 14.68 &    \nodata &      19.21 &      18.44 &    \nodata &    \nodata &    \nodata &    \nodata &    \nodata &                   \nodata &    \nodata &    \nodata &       1.76 & 1/0/0 \\
                 NGC 6809 &          M55 &        CX7 &          19:39:51.19 &          -30:59:01.0 &       M &     NM? &     0.238 &  0.57 &    \nodata &       18.1 &    \nodata &      16.75 &    \nodata &    \nodata &    \nodata &    \nodata &      5.70$\times 10^{30}$ &   0.5-2.5  &    \nodata &       5.59 & 1/0/0 \\
                 NGC 6809 &          M55 &        V64 &          19:39:47.20 &          -30:57:12.3 &       M & \nodata &     5.886 &  0.74 &    \nodata &      17.77 &      17.02 &    \nodata &    \nodata &    \nodata &    \nodata &    \nodata &                   \nodata &    \nodata &    \nodata &      12.95 & 1/0/0 \\
                 NGC 6838 &          M71 &        s02 &          19:53:42.62 &          +18:45:47.2 & \nodata & \nodata &     0.822 &  1.35 &    \nodata &    \nodata &      17.71 &    \nodata &      16.32 &    \nodata &    \nodata &    \nodata &      1.50$\times 10^{31}$ &   0.5-2.5  &    \nodata &    \nodata & 1/0/0 \\
                 NGC 6838 &          M71 &       s19a &          19:53:48.85 &          +18:46:34.0 & \nodata & \nodata &     0.822 &  0.60 &      19.17 &      18.73 &      17.67 &    \nodata &      16.50 &    \nodata &    \nodata &    \nodata &      8.91$\times 10^{31}$ &   0.5-2.5  &    \nodata &    \nodata & 6/0/0 \\
\enddata
\tablenotetext{}{Note: References for values in this table are given in Section~\ref{s:obs}.}
\end{deluxetable}

\setlength{\tabcolsep}{6pt}

\clearpage
\end{landscape}

\end{document}